\documentclass{elsart3-1}

\usepackage{amssymb}
\usepackage{graphicx}
\usepackage{amsmath,amssymb,mathrsfs}
\usepackage[english]{babel}
\usepackage{cite}

\newtheorem{e-proposition}[theorem]{Proposition}

\newtheorem{e-definition}[theorem]{Definition\rm}

\begin{document}

\begin{frontmatter}


\selectlanguage{english}

\title{Simulation of residual oil displacement in a sinusoidal channel with the lattice Boltzmann method}

\footnotetext{This is the pre-peer reviewed version of the following articles: Comptes Rendus M\'ecanique, which has been published in final form at http://www.sciencedirect.com/science/article/pii/S1631072115000893}


\selectlanguage{english}
\author[authorlabel1]{Hiroshi Otomo}\ead{hotomo@exa.com}\author[authorlabel1]{, Hongli Fan}\author[authorlabel2]{, Randy Hazlett}\author[authorlabel1]{, Yong Li,} 
\author[authorlabel1]{Ilya Staroselsky, Raoyang Zhang, Hudong Chen }


\address[authorlabel1]{Exa Corporation, 55 Network Drive, Burlington, Massachusetts 01803, USA}
\address[authorlabel2]{The University of Tulsa, 800 South Tucker Drive, Tulsa, Oklahoma 74104, USA}



\begin{abstract}
 We simulate oil slug displacement in a sinusoidal channel in order to validate computational models and algorithms for multi-component flow.
This case fits in the gap between fully realistic cases characterized by complicated geometry and academic cases with simplistic geometry.
Our computational model is based on the lattice Boltzmann method and allows for variation of physical parameters such as wettability and viscosity. The effect of variation of model parameters is analyzed, in particular via comparison with analytical solutions. We discuss the requirements for accurate solution of the oil slug displacement problem. }

\vskip 0.5\baselineskip
\vskip 0.5\baselineskip
\end{abstract}
\end{frontmatter}


\selectlanguage{english}

%
%

\section{INTRODUCTION}
 Transport properties of immiscible fluids in porous media have been extensively investigated because of their practical and fundamental importance. The challenges here are due to complex multi-component and multi-scale physics, as well as complex geometry. 

One practically important example of such flows is in petroleum reservoir engineering, where the minimum pressure required for removing residual oil is one of the critical rock properties. Due in large part to the limitations of physical experiment, engineers and scientists are increasingly considering numerical simulation, which is very difficult as well. In addition to the usual challenges of complex flow modeling, pores and voids in the rocks have nontrivial topological and geometrical structure. Moreover, the wettability of pore walls that strongly influences the flow through the rock, is hard to take into account in the computational model. In a real rock, the wettability is variable and depends on such factors as mineral composition, microscopic surface roughness, pore shapes, and the adsorption effects \cite{1975_Morrow}.

One of the promising computational approaches for modeling this class of flows is the lattice Boltzmann method (LBM). Its first advantage is its framework that is based on the mesoscopic kinetic theory. Compared with Navier-Stokes - based formalisms, it describes small scale effects more naturally. For example, interfaces between different components are automatically determined once the species' interactions are defined. Detailed modeling of the wall boundaries is also more natural \cite{2013_Colosqui}. Second, the LBM involves only cubic volume lattices that do not adapt to solid boundaries, so that the volume meshing can be made simple and automatic \cite{1998_Chen,2009_Leo,2004_Yanbing,2006_Fan}. Third, the LBM generally has highly parallel computational performance since most of operations are performed locally.

 A number of previous studies that use the LBM for simulation of rock samples and porous media show promising results \cite{2008_Michael,2002_Manwart,1998_Hazlett,2006_Fredrich,1992_Steven,2010_Ramstad,2001_Fang}. However, when a realistic case is simulated it may be difficult to identify the model features that are responsible for deviation of computational results from the experiment and theory.
Therefore, it is desirable to have more basic cases that fit in the gap between fully featured realistic cases and simple academic benchmark tests.
Most of such previously studied simple cases include the capillary rising, the Hagen-Poiseuille flow, the Couette flow, and droplet in free space under specific conditions \cite{2013_Colosqui,2007_Haibo,2008_Haibo,2007_Andreas}. As stated in these papers, essential issues relevant to realistic cases are not accounted for, such as the resolution dependence for complex geometry, transition from steady to unsteady flow regimes, the hysteresis effects, etc. 

In this work, we focus on computation of the minimum pressure required for removing residual oil, which is called the critical pressure, in a sinusoidal channel using a multi-component LBM approach. This geometry can be viewed as a simple prototype of porous media \cite{2001_Fang}. The existence of analytical solution for the critical pressure in this case \cite{1978_Mariano,1979_Mariano,1979_SOO} makes it possible to evaluate the accuracy of predicting the transition from the static to moving slug.
Furthermore, the effects due to resolution, viscosity, and wettability variation upon the quality of numerical results can be evaluated using this prototype model of porous media.

 This paper is organized as follows. In Sec. \ref{sec:FORMALISM}, we review the LBM formalism for multi-component flow. In Sec. \ref{sec:SIMULATION}, we report simulation results. The first case is a two-dimensional droplet in free space, which serves to determine the surface tension. The second case is the two component Hagen-Poiseuille flow, that is used to test the viscous effect. The third case is a two-dimensional slug between flat plates, that is used to define the relation between the contact angle and the corresponding control parameter of the model. After the model parameters are chosen based on these results, the critical pressure for an oil slug in a sinusoidal channel is investigated. In Sec. \ref{sec:SUMMARY}, we summarize the main findings and discuss some potential extensions of this study.

\section{The lattice Boltzmann method for multi-component flow}
\label{sec:FORMALISM}
 Since more than twenty years ago, the LBM has been developed in various ways for simulation of immiscible fluid flows \cite{2014_Haihu}. The LBM model we developed and applied in this study is originated from the well known Shan-Chen model \cite{1993_Xiaowen,1994_Xiaowen}. Combined with other recent LBM advancements \cite{2006_Chen, 2006_Zhang, 2006_Latt, 2006_Xiaowen_JFM, 2012_Qli}, our model provides accurate and stable results, in particular for small viscosity and in arbitrary geometry. The formalism that we use is briefly described below.

The general lattice Boltzmann (LB) equation for multi-component fluid flow, for example that consisting of oil and water, is as follows: 
\begin{equation}
\label{eq:basic_LBM_eq}
 f_{i}^{\alpha} \left( \vec{x}+\vec{c}_{i} \Delta t, t+\Delta t \right) - f_{i}^{\alpha} \left( \vec{x} , t \right)
= \mathcal{C}_{i}^{\: \alpha} + \mathcal{F}_{i}^{\alpha},
\end{equation}
where $f_i^{\alpha}$ is the density distribution function of each fluid component, $\vec{c}_{i}$ is the discrete particle velocity and $\alpha$ is an index for the oil or water component, $\alpha= \left\{ o, w \right\}$. The D3Q19 \cite{1992_Qian} lattice model is adopted here so that the $i$ ranges from 1 to 19. The collision term $\mathcal{C}_{i}^{\: \alpha}$ defines relaxation of particles' distribution functions towards their equilibrium states. $\mathcal{F}_{i}^{\alpha}$ is the term associated with the inter-component interaction force. The most popular and simple form of the collision operator is the BGK operator \cite{1954_Bhantnagar},\cite{1991_Chen,1992_Chen,1992_Qian} with a single relaxation time,
\begin{equation}
\mathcal{C}_{i}^{\: \alpha} = -\frac{1}{\tau^{\alpha}}(f_{i}^{\alpha} - f_{i}^{eq, \alpha}).
\end{equation}
After rearrangement of some terms, the two above equations can be written in the following form,
\begin{equation}
\label{eq:LBM_BGK}
 f_{i}^{\alpha} \left( \vec{x}+\vec{c}_{i} \Delta t, t+\Delta t \right) =
 f_{i}^{eq, \alpha} \left( \rho ^{\alpha}, \vec{u} \right) + \left( 1 -
 \frac{1}{ \tau_{mix} }  \right) f_{i}^{' \alpha} + \mathcal{F}_{i}^{\alpha}.
\end{equation}
Here $\tau_{mix}$ is the "mixed" relaxation time that is related to the kinematic viscosity of the mixture of components:
%
\begin{equation}
\tau_{mix} = \left( \nu_{mix} / T_0  \right) + \frac{1}{2},
\end{equation}
\begin{equation}
\nu_{mix}  = \left( \rho^{o} \nu^{o}+\rho^{w}\nu^{w}) / (\rho^{o}+\rho^{w} \right),
\end{equation}
where $T_0 = 1/3$ is the lattice temperature in D3Q19. The function $f_{i}^{'\alpha}$ is the nonequilibrium particle distribution for each fluid component. It is important that instead of using the standard BGK form $f_{i}^{'\alpha}= f_{i}^{\alpha}-f_{i}^{eq, \alpha}$, a regularized collision procedure is applied in this work in order to calculate $f_{i}^{'\alpha}$,
\begin{equation}
\label{eq:Regualize}
f_{i}^{' \alpha}=\Phi^{\alpha}:\Pi^{\alpha}.
\end{equation}
Here $\Phi$ is a regularization operator that uses Hermite polynomials and $\Pi^{\alpha}$ is the nonequilibrium part of the momentum flux. The basic concept of regularized collision procedure can be found in \cite{2006_Chen, 2006_Zhang, 2006_Latt, 2006_Xiaowen_JFM,1997_Chen}. $f_{i}^{eq}$ is the equilibrium distribution function with the third order expansion in $\vec{u}$,
\begin{equation}
\label{eq:feq}
f_{i}^{eq, \alpha}(\rho^{\alpha}, \vec{u}) = \rho^{\alpha} w_{i} \left[ 1 + \frac{\vec{c}_{i} \cdot \vec{u}}{T_0}  + \frac{\left( \vec{c}_{i} \cdot \vec{u} \right)^2}{2T_0^2} - \frac{\vec{u}^2}{2T_0} +  \frac{\left( \vec{c}_{i} \cdot \vec{u} \right)^3}{6T_0^3} -  \frac{\vec{c}_{i} \cdot \vec{u}}{2T_0^2}\vec{u}^2 \right].
\end{equation}
 
For multi-component flows, a non-local interaction force between respective components should be considered in addition to pure molecular collision. Given component densities $\rho^{\alpha}$ and $\rho^{\beta}$, the interaction force $\vec{F}^{\alpha, \beta}$ acting on the component $\alpha$ due to existence of the component $\beta$ is defined according to \cite{1993_Xiaowen,1994_Xiaowen} as,
\begin{equation}
\label{eq:comp_interaction}
\vec{F}^{\alpha, \beta} \left( \vec{x} \right) = G \rho^{\alpha} \left( \vec{x} \right) \sum_{i} w_{i} \vec{c}_{i} \rho^{\beta} \left(  \vec{x}+ \vec{c}_{i} \Delta t \right) .
\end{equation} 
Here, $G$ denotes a parameter which defines interaction strength and $w_{i}$ is the isotropic weight in D3Q19 \cite{1992_Qian}. Following \cite{2012_Qli}, the interaction force is integrated in $f_i^{eq,\alpha}$ as,
\begin{equation}
\label{eq:collision_term}
f_{i}^{eq, \alpha } (\rho^{\alpha}, \vec{u}) = f_{i}^{eq, \alpha} ( \rho^{\alpha} , \vec{u}_{mix} + \vec{F}^{\alpha, \beta} / \rho^{\alpha}),
\end{equation}
where $\vec{u}_{mix}$ is a mixture velocity,
\begin{equation}
\vec{u}_{mix} = \frac{\vec{u}^{o} \rho^{o} + \vec{u}^{w} \rho^{w}}{\rho^{o} + \rho^{w}}.
\end{equation}

In an immiscible two component system such as oil and water, the inter-component interaction force is repulsive. $\tau_{mix}$ becomes equal to the relaxation time of the respective single component in the regions away from the interface. It can be readily shown that the correspondent macroscopic equation in this framework is the Navier-Stokes equation for a mixture of ideal gases with a repulsive inter-component force \cite{1995_Xiaowen,1996_Xiaowen}. As a result, an equation of state for the nonideal gas is obtained.

In this study, we utilize the volumetric boundary condition for arbitrary geometry proposed originally by Chen et al in 1998 \cite{1998_Chen,2009_Leo,2004_Yanbing,2006_Fan}. In brief, the curved solid surface is discretized into piece-wise linear surface facets in two dimensions and triangular polygons in three dimensions. During each fluid dynamics calculation, the facets/polygons first collect incoming particles from the neighboring cells. Then the outgoing particles are calculated by following the desired collision rules on the surface, for example, reversing particles to realize no-slip/bounce back boundary condition. Finally, the outgoing particles are scattered back into neighboring cells in a volumetric way such that conservation laws are enforced locally. Full algorithm details can be found in \cite{1998_Chen}. It is worth pointing out that this approach does satisfy the local detailed balance and is the only scheme, as far as we know, that can accurately realize the frictionless wall in an arbitrary geometry. Surface wettability in a multi-component flow LBM model is usually based on introducing wall potentials \cite{2007_Haibo,2008_Haibo,2007_Andreas}. In our case, the wall potentials correspond to $\rho_s^{o}$ and $\rho_s^{w}$ defined on the solid surface, which control the surface contact angle. In calculation of the interaction force in near wall cells, the wall potentials are applied in a consistent volumetric way \cite{1998_Chen} to replace the missing neighbor cells in Eq. (\ref{eq:comp_interaction}) such that the accuracy of surface interaction force can be maintained in complex geometries.

\section{SIMULATION}
\label{sec:SIMULATION}

Before focusing on the critical pressure, systematic studies on surface tension, viscous effects, and the contact angle in our LB model were carried out first. All the physical quantities in this paper are in lattice units, and the discrete lattice time and space increments are $\Delta x = \Delta t = 1$.

\subsection{Estimation of the surface tension with a two-dimensional droplet in free space}
\label{subsec:surface_tension}
The surface tension $\sigma$ is measured with a static two-dimensional droplet in free space. The pressure and surface tension are related via the Laplace law:
\begin{equation}
\label{Young-Laplace_equation}
 dP=\frac{\sigma}{R} \: ,
 \end{equation}
where $dP$ is pressure difference across the droplet interface and $R$ is the droplet radius. The simulation domain is a square with the resolution of 2.5 times the droplet diameter. The initial densities of both components are 0.22. Four sets of cases with different initial droplet diameters $\left\{ 16, 24, 32, 48 \right\}$ and different relaxation times $\left\{ \tau_{1}=1.0 , \tau_{2}=1.0 \right\}$ , $\left\{ \tau_{1}=0.55 , \tau_{2}=1.5 \right\}$ and $ \left\{ \tau_{1}=1.5 , \: \tau_{2}=0.55 \right\} $  are tested. Here the subscripts $1$ and $2$ denote quantities inside and outside the droplet, respectively. The last two sets of viscosity ratios achieve the value of 20, which matches the viscosity ratio between oil and water. Such high viscosity ratio is hard to reproduce using alternative LBM schemes.

\begin{figure}[htbp]
  \begin{center}
          \includegraphics[clip, width=8 cm]{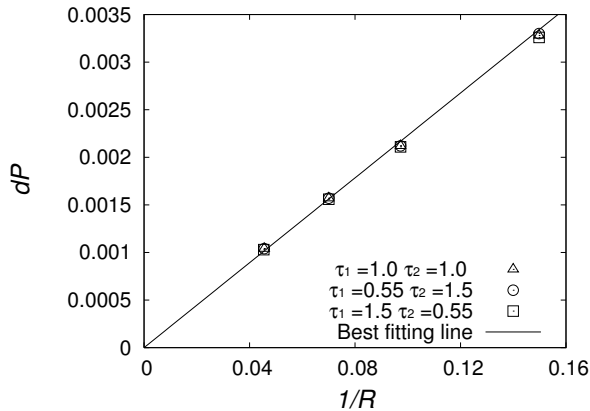}
	  \caption{Pressure difference across droplet interface as a function of the inverse droplet radius. The results from three sets of relaxation times are plotted. The slope of the best fit line defines the surface tension in terms of the Young-Laplace equation, Eqs. (\ref{Young-Laplace_equation}). }
	   \label{fig:laplace}
  \end{center}
\end{figure}

To minimize the numerical error caused by finite interface thickness, density profiles $\rho (x_{i})$ (where $i$ indexes the discrete coordinates) along the central vertical and horizontal lines are chosen. They are fitted by hyperbolic tangent functions: 
\begin{equation}
\label{den-fitting}
\rho (x_i) = C_{1} \tanh \left( C_{2} \left( x_i - C_{3}  \right) \right) + C_{4} \:.
\end{equation}
$C_{1}$, $C_{2}$, $C_{3}$ and $C_{4}$ are fitting constants , .e.g, $C_{3}$ is the position of the middle point of the interface between oil and water. The radius $R$ can be determined with little uncertainty by measuring the distance between the two interfaces along the chosen lines. On the other hand, pressure values at the droplet center and a point far away from the droplet are measured and their difference gives the $dP$ value. The results for all cases are shown in Fig. \ref{fig:laplace}, where the solid line fit over all the data sets. According to Eq. (\ref{Young-Laplace_equation}), the slope of the solid line is indeed the surface tension $\sigma$. Observe that our procedure self-consistently returns a value of $\sigma$ which shows neither resolution nor viscosity dependence. Note also that our model can achieve accurate and stable results even when $\tau$ is small.

As a result, we obtain 
\begin{equation}
\sigma = \: 2.51e \mathchar`- 2 \: \pm \: 0.02e\mathchar`- 2 \: .
\end{equation}
The numerical uncertainty mainly comes from measurement fitting of dP vs 1/R and the density profile, as well as spatial fluctuation in the bulk regions. The uncertainty due to pressure fluctuation in the interface region is insignificant.

\subsection{Evaluation of the viscous effects with a two-component Hagen-Poiseuille flow}
\label{subsec:two-comp_Poiseuille}
The viscous effects are studied on the test case of the two-component Hagen-Poiseuille flow. One component is oil with $\tau_{o}=1.5$ and the other is water with $\tau_{w}=0.55$. Their viscosity ratio is 20 representing a realistic situation. The channel height is 40 and the driving force is chosen as $1.6e\mathchar`- 6 $. The density ratio between two components is 1. Two cases with different initial component distributions are tested. In the first case, oil occupies the center region of the channel and water flows along the walls. In the second case, their initial positions are exchanged.  

\begin{figure*}[htbp]
  \begin{center}
    \begin{tabular}{c}
      \begin{minipage}{0.5\hsize}
        \begin{center}
          \includegraphics[clip, width=7.5 cm]{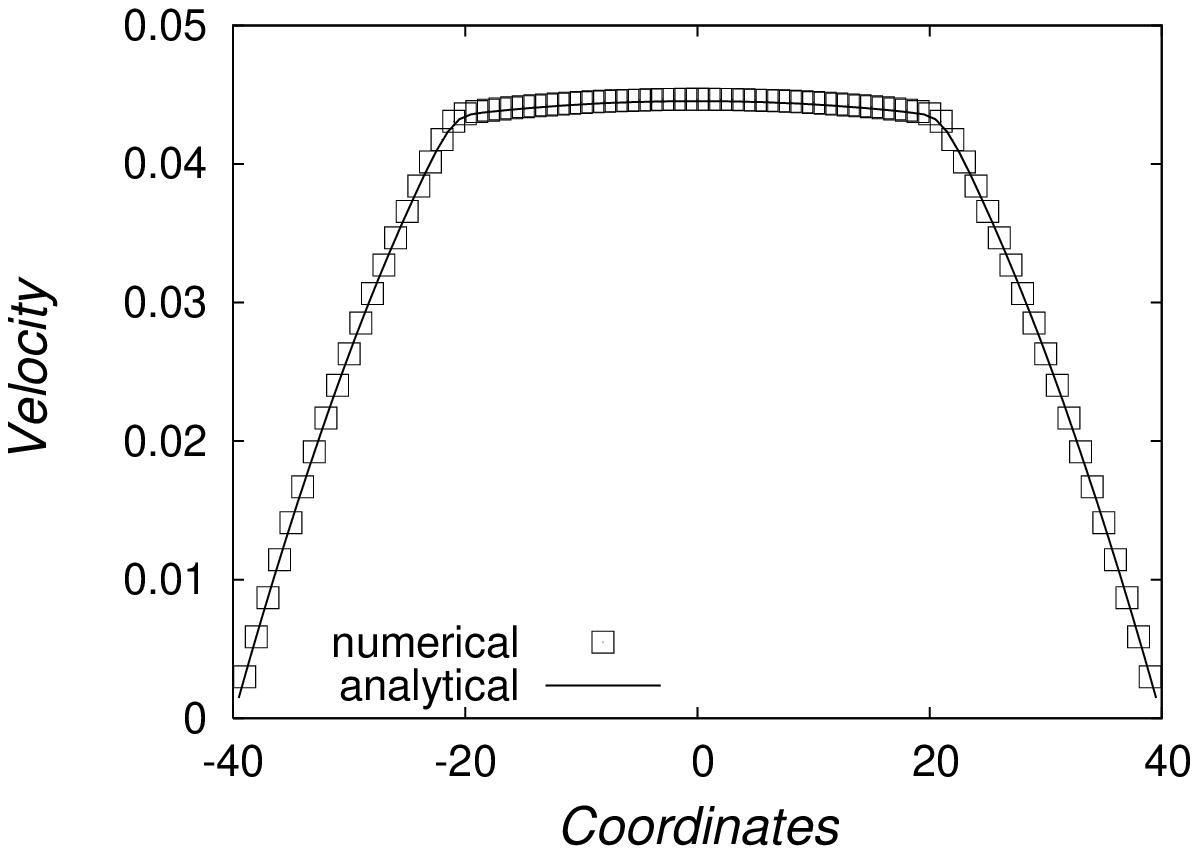}
        \end{center}
      \end{minipage}
      \begin{minipage}{0.5\hsize}
        \begin{center}
          \includegraphics[clip, width=7.5 cm]{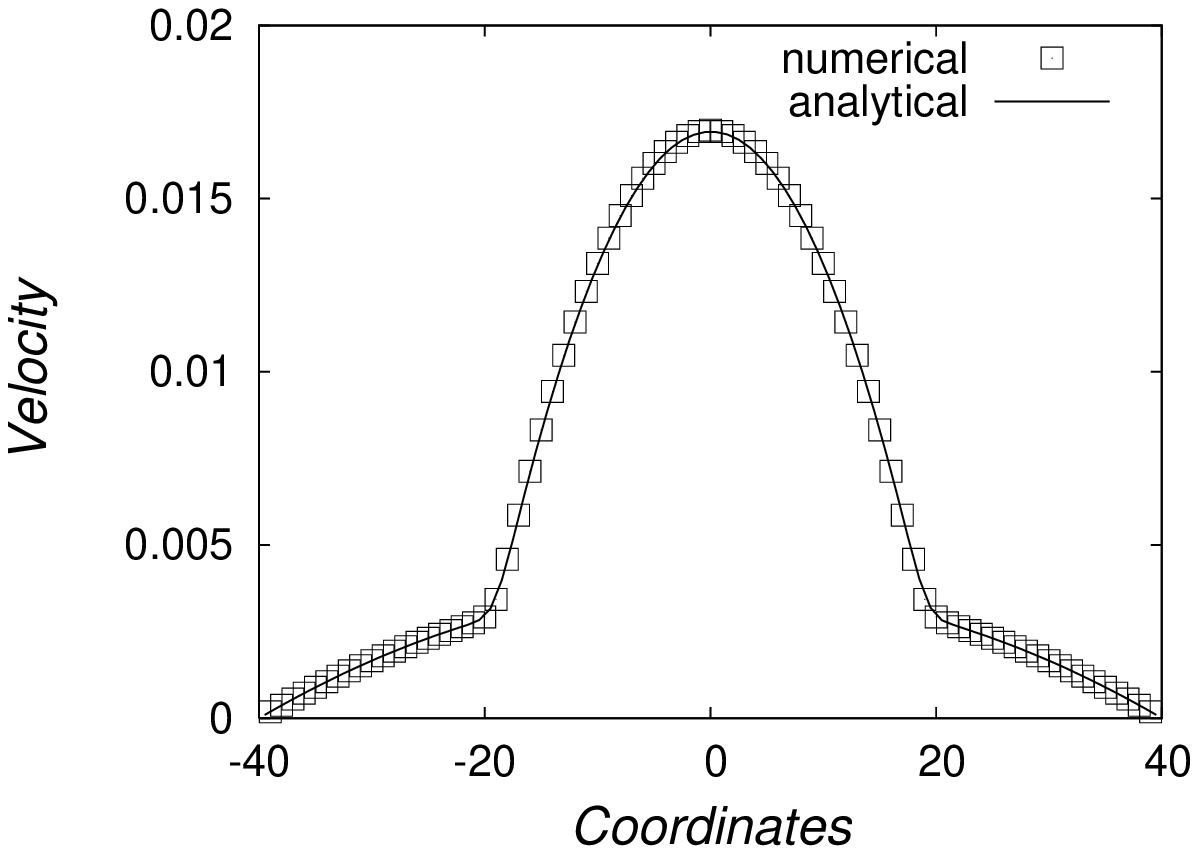}
        \end{center}
      \end{minipage}
    \end{tabular}
    \caption{Velocity profiles across the channel for the Hagen-Poiseuille flow of oil and water simulated with $\tau_{o}=1.5$ and $\tau_{w}=0.55$. Symbols are numerical results and solid lines are analytical solutions. Oil is in the central region and water is along the walls (Left). Oil is along the walls and water is in the center (Right).}
    \label{fig:two-comp-Poiseuille}
  \end{center}
\end{figure*}

The simulation results are presented in Fig. \ref{fig:two-comp-Poiseuille} together with exact analytical solutions. The very good match demonstrates accuracy and robustness of our two component LB model.

\subsection{Estimation of the contact angle based on the test case of a slug between flat plates}
\label{subsec:contact_angle}
The relation between the contact angle $\theta$ and its control parameter, wall potential, is studied using a two-dimensional slug between two parallel plates. The distance $H$ between the plates is 32 and the initial density of both components is 0.22. Two sets of relaxation times are tested, one is $ \tau_{w}=1.0 , \: \tau_{o}=1.0 $ and the other is $\tau_{w}=0.55 , \: \tau_{o}=1.5 $, corresponding to the viscosity ratio of 1 and 20, respectively.

\begin{figure}[htbp]
  \begin{center}
          \includegraphics[scale=0.02]{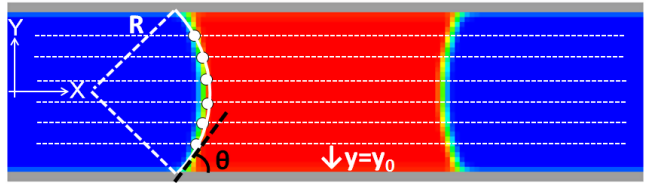}
	  \caption{ Static contact angle detection. The color contours shows water density distribution.}
	   \label{fig:contact_angle_estimation}
  \end{center}
\end{figure}

The choice of wall potential values in two-component flow simulations for a particular contact angle is not unique. The wall potential values are positive in our practice for the purpose of better stability and accuracy. Fig. \ref{fig:contact_angle_estimation} shows the contact angle detection in a schematic way. First, the interfacial points (solid white circles) are identified on each interface by the line fitting using hyperbolic tangent test functions (Eq. (\ref{den-fitting})) along the x-coordinate. Then a two dimensional circle is constructed in order to fit these interfacial points. The contact angle $\theta$ can be obtained by measuring the slope, $dy/dx$, of the tangential line across the contact point:
\begin{equation}
 \label{eq:contact_angle}
 \theta=\arctan \left( \left( \frac{dy}{dx} \right)_{y=y_0} \right)=\arctan \left( \frac{\sqrt{R^2-y_{0}^2}}{y_0} \right) \: ,
\end{equation}
where $y_0$ is the half channel width and $R$ is the radius of the fitted circle.

Fig. \ref{fig:contact_angle} shows the contact angle as a function of the wall potential value. The presented data also includes the dependence upon the offset of solid boundary from the lattice and upon viscosity. Offset establishes the physical location of the solid-fluid interface at sub-lattice dimensions. The wall potential is normalized with fluid component density
0.22. Different viscosity combinations were tried and only two representative sets, $\left\{ \tau_{w}=1.0, \tau_{o}=1.0 \right\}$ and $\left\{ \tau_{w}=0.55, \tau_{o}=1.5 \right\} $, are shown. Note that the contact angle in our model shows only very small dependence upon both the boundary offset and viscosity, which again demonstrates the accuracy and robustness of the scheme. We believe that the small deviations observed in Fig. \ref{fig:contact_angle} are mainly caused by numerical smearing in the near wall regions at finite resolution. For the study of critical pressure presented below, however, this small variability can not be totally neglected. 

In theory, the angles $\theta$ presented in Fig. \ref{fig:contact_angle} should change to $180-\theta$ when the substance is changed from oil to water and vice versa, so that an ideal profile should have a center of symmetry relative to the point where it intersects the vertical line originated at zero of wall potential. While this is almost true, a slight asymmetry observed in Fig. \ref{fig:contact_angle} is believed to be caused by numerical uncertainty in the surface tension measurement discussed earlier. 

Resolution independence of these results has been confirmed for $H=\left\{ 16, 32, 64 \right\}$. Results for two sets of the normalized wall potential, $\left\{ \rho_s^{w}= 0, \: \rho_s^{o}=0.44 \right\}$ for the oil-wet case and $\left\{ \rho_s^{w}=0.44, \: \rho_s^{o}=0 \right\}$ for the water-wet case are shown in Table \ref{tab:contact_angle_pickup}. We conclude that the contact angle is 141.5 $\pm$ 3.5 degree for the oil-wet case and 41.5 $\pm$ 2.6 degree for the water-wet case. These two sets of the contact angle are used for our study of the critical pressure below. 

\begin{figure*}[b]
  \begin{center}
    \begin{tabular}{c}
      \begin{minipage}{0.5\hsize}
        \begin{center}
          \includegraphics[clip, width=8 cm]{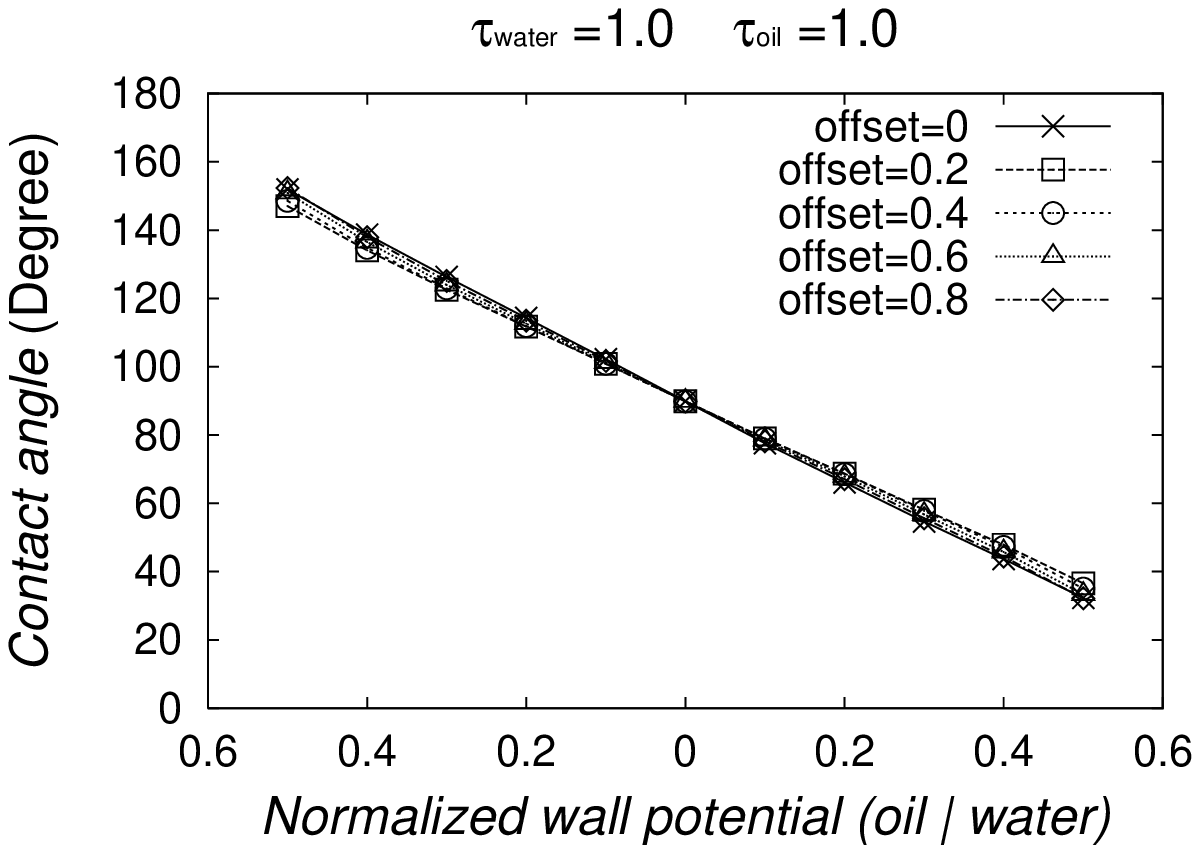}
        \end{center}
      \end{minipage}
      \begin{minipage}{0.5\hsize}
        \begin{center}
          \includegraphics[clip, width=8 cm]{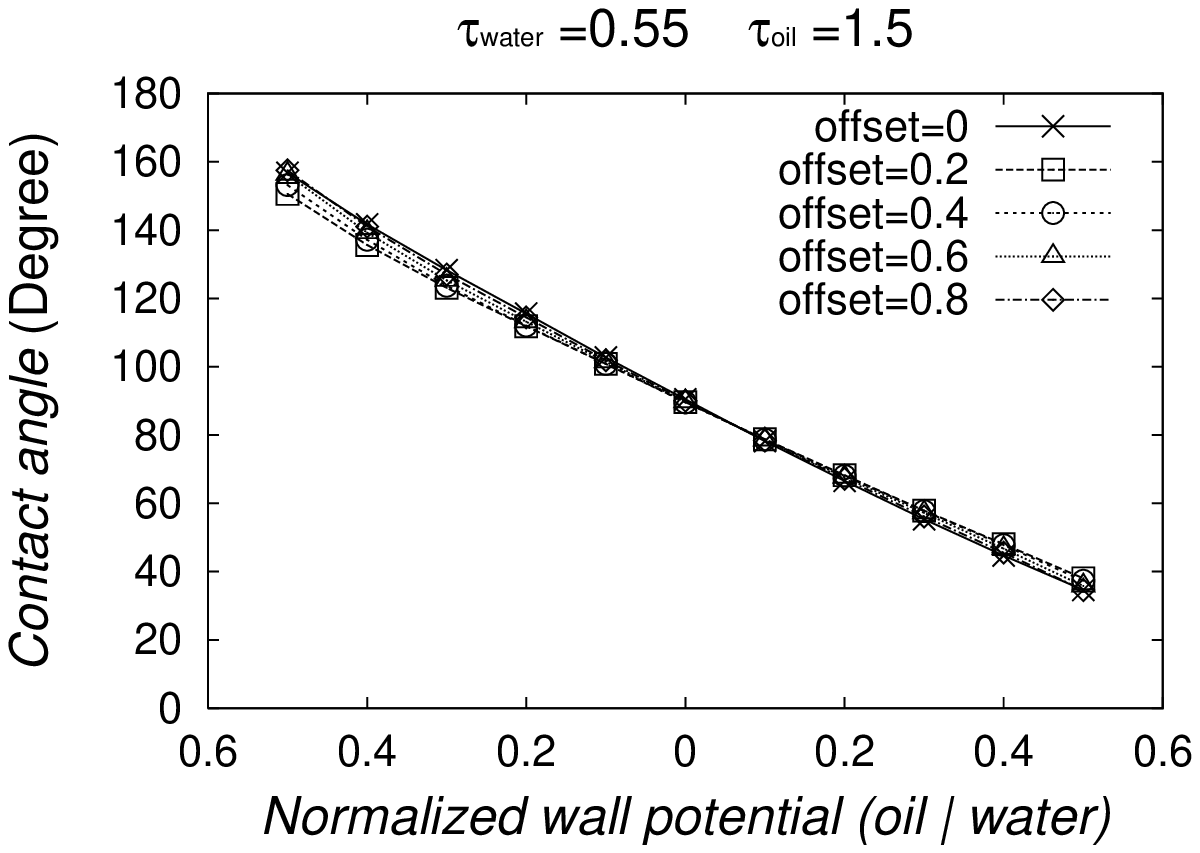}
        \end{center}
      \end{minipage}
    \end{tabular}
    \caption{The static contact angle as a function of normalized wall potential with offset boundaries. The horizontal axis represent the wall potential for oil on the left side of central vertical axis and wall potential for water on the right side. Wall potential is normalized by typical density value 0.22. The relaxation times are $\tau_{w}$=1.0, $\tau_{o}$=1.0 (Left) and $\tau_{w}$=0.55, $\tau_{o}$=1.5(Right).}
	   \label{fig:contact_angle}
    \label{fig:contact_angle}
  \end{center}
\end{figure*}

\subsection{Critical pressure for an oil slug in a sinusoidal channel}

The critical pressure for oil slug displacement in a sinusoidal channel is simulated with variable wettability, resolution and viscosity. When a static slug is subject by pressure force, it mainly reacts with the surface tension, i.e. the capillary force. For the quantitative assessment of critical pressure, a useful dimensionless number can be defined as the ratio of the pressure and capillary forces. In the context of our simulation, note that in order to minimize the artificial compressibility effects, we apply a constant driving body force $g$ instead of the regular pressure. The dimensionless ratio of forces (usually referred to as the Bond number $Bo$) can be written as 
\begin{equation}
 \label{Bo_number}
 \frac{\rm Pressure \ force}{\rm Capillary \ force} = \frac{\Delta P D^{2}}{\sigma D } \sim \frac{\rho g L D }{\sigma }.
\end{equation}
For the sinusoidal channel studied here, $D$ is defined as the radius at the neck and $L$ as the channel length. A slug starts to move when the Bond number exceeds a certain critical value. Analogous to friction, we recognize that the forces required to initiate motion and to sustain it can be different. This may be reconciled with static and dynamic contact angles, something not entertained here. Once the geometry and fluid properties are determined, only $g$ is variable in Eq. (\ref{Bo_number}). Therefore, in order to detect the critical pressure, we change the $g$ value periodically and look for the critical Bond number for oil slug displacement. 

\begin{figure*}[htbp]
  \begin{center}
    \begin{tabular}{c}
      \begin{minipage}{0.5\hsize}
        \begin{center}
          \includegraphics[clip, width=8 cm]{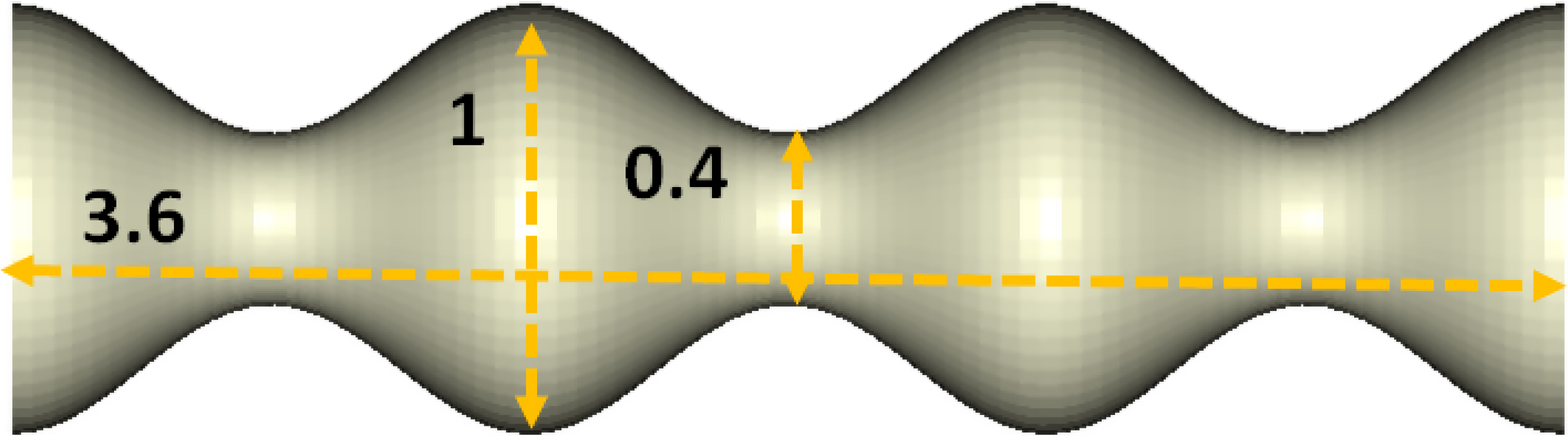}
        \end{center}
      \end{minipage}
      \begin{minipage}{0.5\hsize}
        \begin{center}
          \includegraphics[clip, width=8 cm]{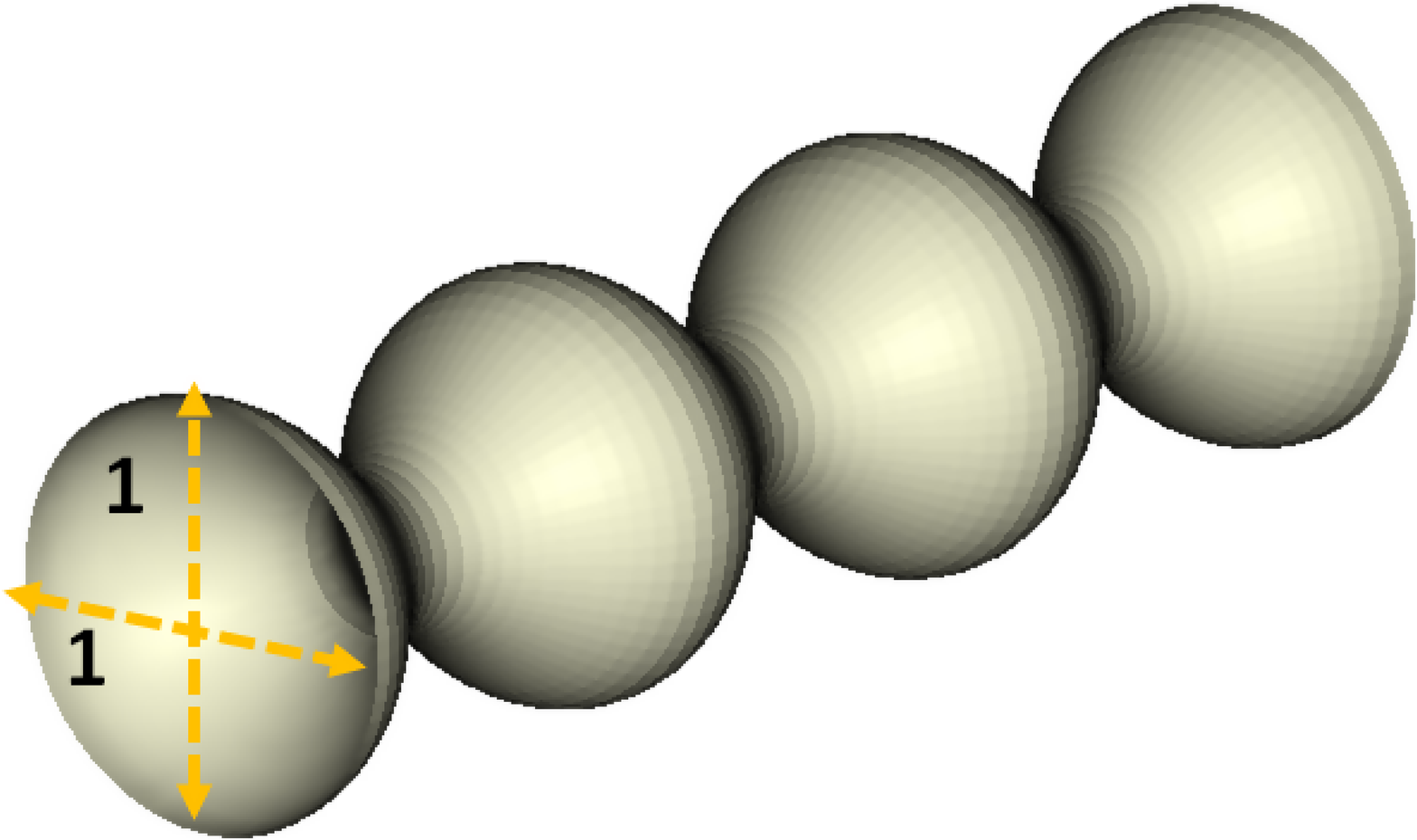}
        \end{center}
      \end{minipage}
    \end{tabular}
    \caption{Geometry of the sinusoidal channel. In the units of the maximum channel height, the channel length is 18 which include three cycles of the sine curve, and the channel height at the neck is 0.4. All the cross sectional areas are circles. The wall surface contains 38,570 mesh elements. }
    \label{fig:sinusoidal_geom}
  \end{center}
\end{figure*}

\begin{figure}[htbp]
  \begin{center}
          \includegraphics[clip, width=8 cm]{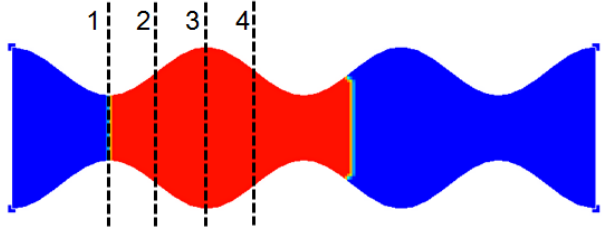}
	  \caption{ Different initial slug positions. The left edges of initial slugs are depicted with dotted lines and set in four ways. The color contours represent the initial oil density distribution for the oil-wet case in the position 1. }
	   \label{fig:initial_slug_position}
  \end{center}
\end{figure}

The geometry of the sinusoidal channel is shown in Fig. \ref{fig:sinusoidal_geom}. The channel wall is composed with 38570 surface elements for sufficient discrete surface mesh quality. The wall is set as no-slip with the ability to specify preferential fluid interactions, i.e. wettability. The wall potential is set to achieve the contact angle $\theta \approx 40$ degrees for the water-wet, $\theta$ $\approx$ 90 degrees for the intermediate(neutral) and $\theta$ $\approx$ 140 degrees for the oil-wet cases.  Three resolutions across the neck corresponding to $D= \left\{ 6, 8, 12 \right\} $ are applied and the initial densities of both oil and water are set to 0.22. The wavelength and amplitude representing a pore body are held constant to mimic the physical common association with grain size in clean, well-sorted sandstone. The range of pore neck investigated also covers a reasonable spread of pore body-to-pore throat ratio encountered unconsolidated packs of spheres or sieved sand \cite{Payatakes_1973}. The periodic boundary is enforced in the flow direction. In order to check the influence of initial condition, four initial slug positions shown in Fig. \ref{fig:initial_slug_position} were tried for each parameter set. In the current validation cases, the hysteresis effects are not taken into account explicitly and the advancing and receding dynamic contact angles are controlled by the same wall potential.


\begin{figure*}[htbp]
  \begin{center}
    \begin{tabular}{c}
      \begin{minipage}{0.5\hsize}
        \begin{center}
          \includegraphics[clip, width=7.5 cm]{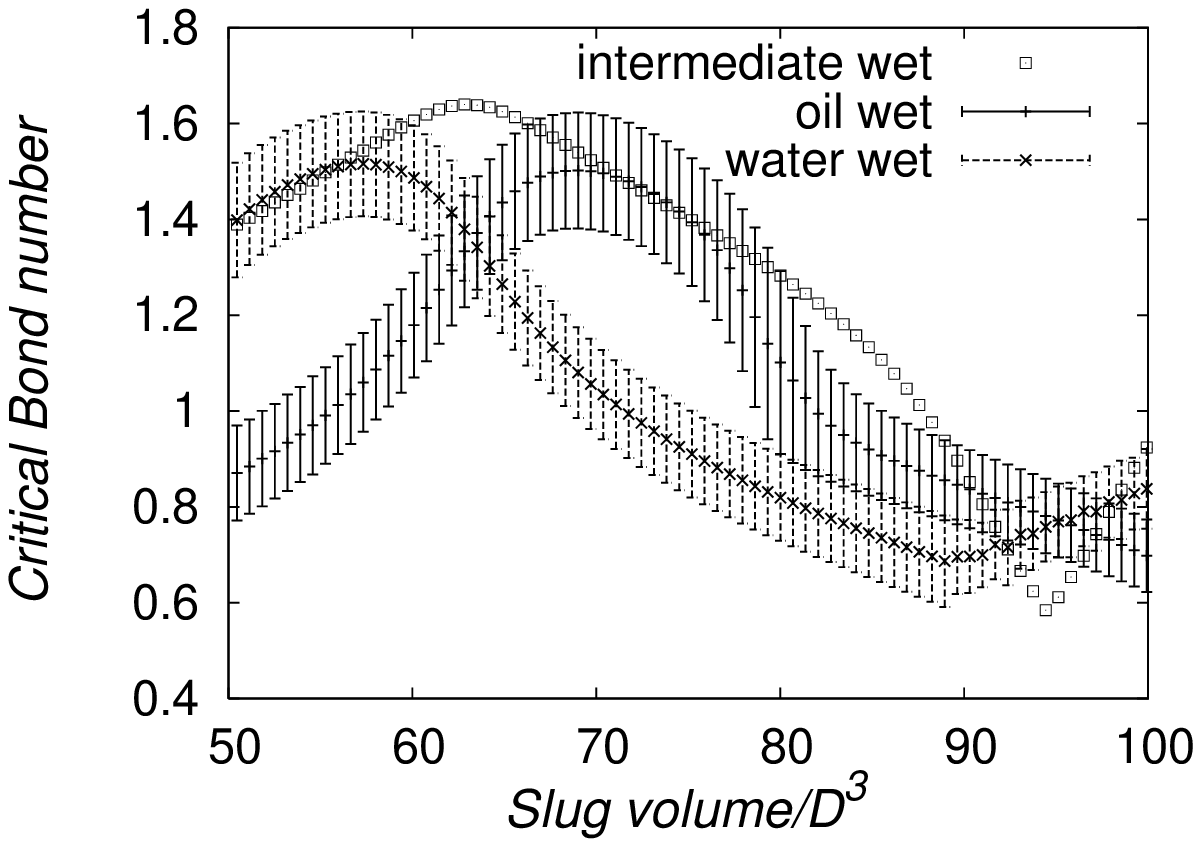}  
        \end{center}
	\caption{Analytical solutions for the critical Bond number as a function of the slug volume non-dimensionalized by $D^{3}$. Error bar is derived from the contact angle variance caused by lattice non-alignment shown in Table. \ref{tab:contact_angle_pickup}.}
	\label{fig:Bo_slugvol_dep}
      \end{minipage}
      \begin{minipage}{0.5\hsize}
        \begin{center}
          \includegraphics[clip, width=7.5 cm]{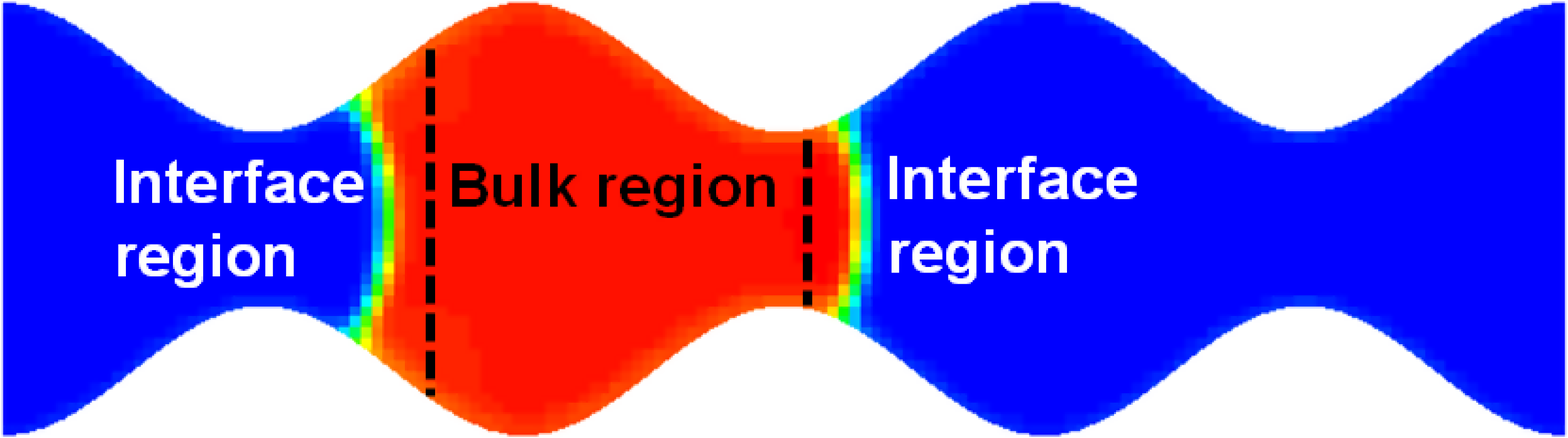}  
        \end{center}
	\caption{Slug volume detection. Around the contact points and interfaces, vertical lines divide the oil slug into the oil bulk region and interface regions.}
	\label{fig:volume_detection}  
      \end{minipage}
    \end{tabular}
  \end{center}
\end{figure*}

The analytical solutions for critical pressure for different setups were calculated by Oh and Slattery (in \cite{1979_SOO}). The availability of these analytical results is an important motivating factor for the present study. According to Eqs (3) in \cite{1979_SOO}, the critical Bond number is written as
\begin{equation}
Bo_{crit}=-2D \left( H_{wo} + H_{ow} \right),
\end{equation}
where $H_{wo}$ and $H_{ow}$ are curvatures of the receding and advancing interfaces estimated by (A-9) and (A-10) in \cite{1979_SOO}. They are determined by the channel geometry, wettability of the wall, and the slug volume (contact points). In Fig. \ref{fig:Bo_slugvol_dep}, the dependence of analytical solutions upon the oil volume is plotted. The targeted slug volume in our simulation is chosen in Table. \ref{table:slug_volume} such that the critical $Bo$ number has the least variability with respect to the slug volume. Due to the limitations of both physical and numerical models, there is always a small artificial oil/water residue in the water/oil regions. Therefore an appropriate volume detection approach needs to be formulated. In our simulation, for the numerical measurement of the slug volume, the oil slug region is separated to the bulk region and interface regions using positions of interfaces and contact points as shown in Fig. \ref{fig:volume_detection}. In bulk region, lattice volume is simply summed. In interface regions it is summed only where the oil density is more than 90 percent of the total density value.

The first result presented here corresponds to the case with $\theta \approx 90$ (neutral), $D=8$, and $\tau=1$ for both components. During the initial 40,000 time steps, the slug is not subject to external body force but does change shape subject to interfacial tension and wettability considerations. Then the body force is turned on with $g=1.47e-4$ (corresponding to $Bo = 1.48$) and is ramped up by 1.98e-5 ($\Delta Bo = 0.02$) every 40,000 time steps. The center of oil mass is calculated in order to track down the slug position.

\begin{figure}[htbp]
  \begin{center}
          \includegraphics[clip, width=8 cm]{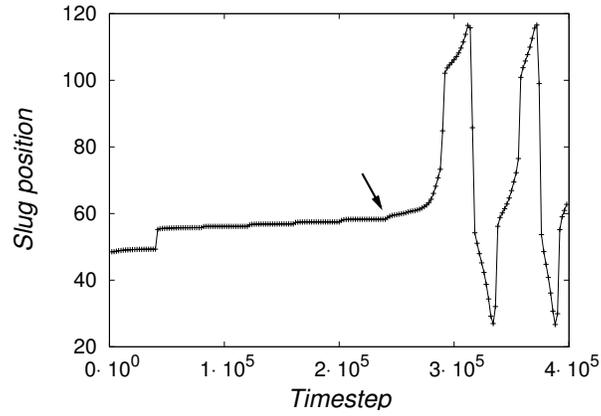}
	  \caption{ Time history of a slug position (the center of oil mass) in the intermediate-wet case with $D=8$ and $\tau_{w}=1,\tau_{o}=1$. }
	   \label{fig:time_history_of_slug_position}
  \end{center}
\end{figure}

Its time history is presented in Fig. \ref{fig:time_history_of_slug_position}. For the initial 40,000 timesteps, the slug is at the initial position due to zero external force.

\begin{figure*}[htbp]
  \begin{center}
    \begin{tabular}{c}
      \begin{minipage}{0.5\hsize}
        \begin{center}
          \includegraphics[clip, width=7.5 cm]{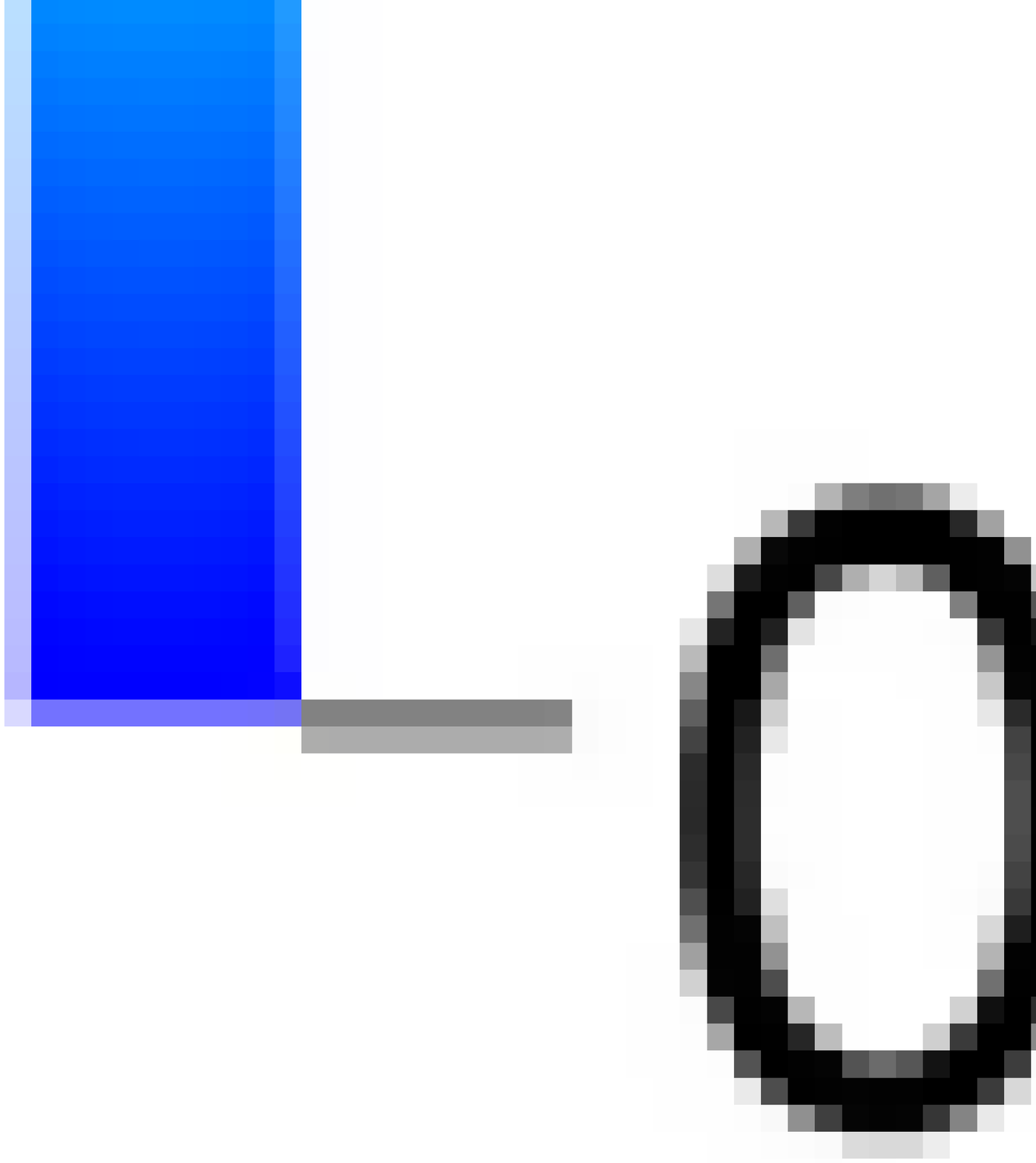}
        \end{center}
      \end{minipage}
      \begin{minipage}{0.5\hsize}
        \begin{center}
          \includegraphics[clip, width=7.5 cm]{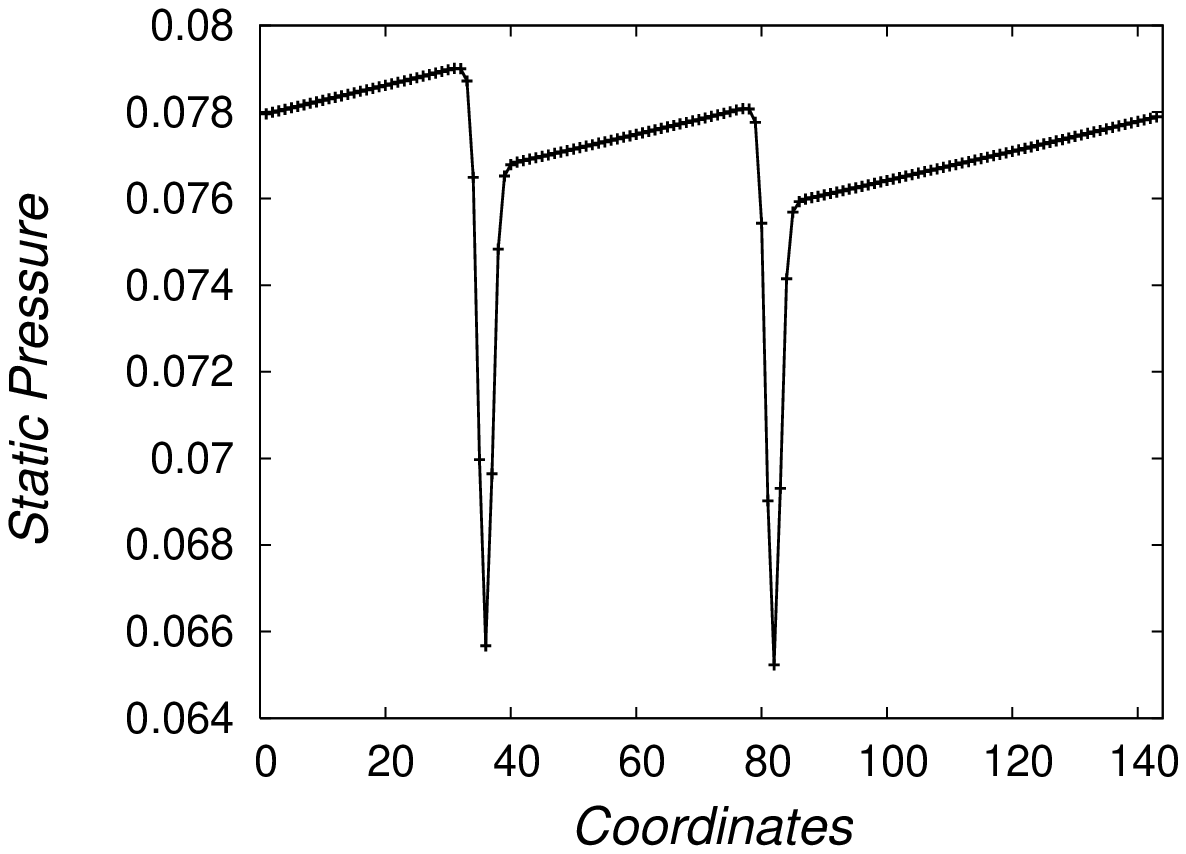}
        \end{center}
      \end{minipage}
    \end{tabular}
    \caption{Color contours of static pressure (Left) and pressure profile along the central horizontal line (Right) at 200,000 time steps for the intermediate-wet case. The shaded area on the left is the oil slug region where the density exceeds 0.11.}
    \label{fig:prs_image_intermediate_tau1_pos2}
  \end{center}
\end{figure*}

\begin{figure}[htbp]
  \begin{center}
          \includegraphics[clip, width=8 cm]{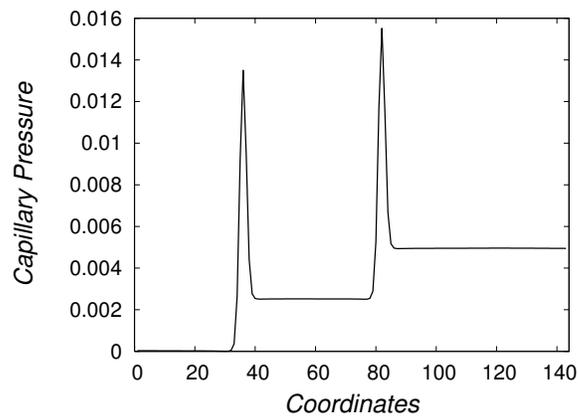}
	  \caption{ Capillary pressure calculated based on the results from Fig. \ref{fig:prs_image_intermediate_tau1_pos2}. }
	   \label{fig:Cap_prs_plot_intermediate_tau1_pos2}
  \end{center}
\end{figure}

Then the slug's center position is slightly shifted due to the surface shape change needed to balance the capillary force and the external body force. When the external body force takes over the capillary force around 240,000 time steps, the slug is pushed enough to move and eventually squeeze through the channel neck quickly. Fig. \ref{fig:prs_image_intermediate_tau1_pos2} shows the pressure distribution at 200,000 time steps at $g=1.55e-4$. On the left side of Fig. \ref{fig:prs_image_intermediate_tau1_pos2}, the slug area is shaded and color contours represent the static pressure. The pressure cusps such as observed at the interface have been discussed in \cite{2008_Xiaowen}. Here we focus on the pressure difference in the separate bulk regions. According to the Laplace law, the pressure should decrease from left to right side across both interfaces. The pressure profile $P \left( x \right)$ in the graph on the right side of Fig. \ref{fig:prs_image_intermediate_tau1_pos2} does show the correct trend. To further demonstrate this effect, the hydrostatic pressure due to the external body force is removed,
 \begin{equation}
 \label{plot_value}
 P \left( 0 \right)- \left[ P  \left( x \right) - \rho \cdot g \cdot x \right]
\end{equation}
and the result is shown in Fig \ref{fig:Cap_prs_plot_intermediate_tau1_pos2}. The difference between the left and right channel ends is the total capillary pressure that is balanced by $\rho g  x = 0.22 \cdot 1.55e \mathchar`- 4 \cdot 144 = 0.00491$. On the other hand, by considering that $P  \left( x \right) - \rho \cdot g \cdot x$ is equivalent to the inherent pressure distribution and ignoring the behaviors around interfaces, one can see the benefit of using the homogeneous body force scheme. Indeed the comparison between Fig. \ref{fig:prs_image_intermediate_tau1_pos2} and Fig. \ref{fig:Cap_prs_plot_intermediate_tau1_pos2} shows that the pressure range is reduced overall and as a result the artificial compressibility is reduced. 


At the time step of about 240,000 indicated by the arrow in
Fig. \ref{fig:time_history_of_slug_position}, the driving force $g=1.57e
\mathchar`- 4$ exceeds the capillary force and pushes the slug enough to start to
move. Therefore the critical driving force $g_{crit}$ is between 1.55e-4 and
1.57e-4. The resulting Bond number is
\begin{equation}
 \label{Bo_number_example}
  \frac{\rho g L D }{ \sigma }=\frac{0.22 \cdot (1.55e \mathchar`- 4 \sim  1.57e \mathchar`- 4) \cdot 144 \cdot 8}{ 2.51e \mathchar`- 2 \: \pm \: 0.02e\mathchar`- 2 }=1.56 \sim 1.59.
\end{equation}
The oil slug volume measured numerically using the scheme described above is 58.2 at 200,000 time steps. Therefore, based on Fig. \ref{fig:Bo_slugvol_dep}, the analytical critical $Bo$ number is 1.57.

Similar simulations and analysis are performed for the other three initial slug
positions. In all cases, the same $ g_{crit} $ is achieved and the oil
slug volumes in the static state are detected as 57.9, 58.1, 58.1
respectively. We can summarize that the computationally achieved critical $Bo$ number 1.56 $\sim$
1.59 is consistent with the analytical prediction. The numerical and analytical results deviate within $1.3 \%$ with no dependence upon the slug initial position.

\subsubsection{The wettability factor}
\label{subsubsec:wettability_depenency}
Now we focus on the oil-wet ($\theta \approx 140$ degree) and water-wet ($\theta \approx 40$ degree) cases. The normalized wall potential has the same value as in the test from Table \ref{tab:contact_angle_pickup}. Similar to the first neutral case, the simulation resolution is $D=8$ and $\tau=1$ for both components. The increment of ramping up $g$ is adjusted similarly such that $\Delta Bo = 0.02$.

The simulation results are shown in the Table. \ref{tab:crit_force_Bond_water-wet_oil-wet}. The numerically measured slug volume is listed in the third column, and the resulting $Bo$ is in the fourth column. Without considering the difference between the static and dynamic contact angles, the analytical prediction is listed in the sixth column. The difference between the simulation and analytical solutions is about $6\%$. When we consider the $\Delta \theta$ contribution to the advancing and receding contact angles in the calculation of the analytical solution, the highest $Bo$ number is presented in the fifth column and the comparison then becomes very good $\sim 1\%$. Here $\Delta \theta$ is the contact angle variance presented in Fig. \ref{fig:contact_angle}. Therefore it seems important to take the hysteresis effect into account for quantitatively accurate numerical modeling and simulation. In addition, we note that, although the lattice dependence of contact angle is small, it needs to be considered explicitly for the purpose of accuracy as well.

\begin{figure*}[htbp]
  \begin{center}
    \begin{tabular}{c}
      \begin{minipage}{0.5\hsize}
        \begin{center}
          \includegraphics[clip, width=7.5 cm]{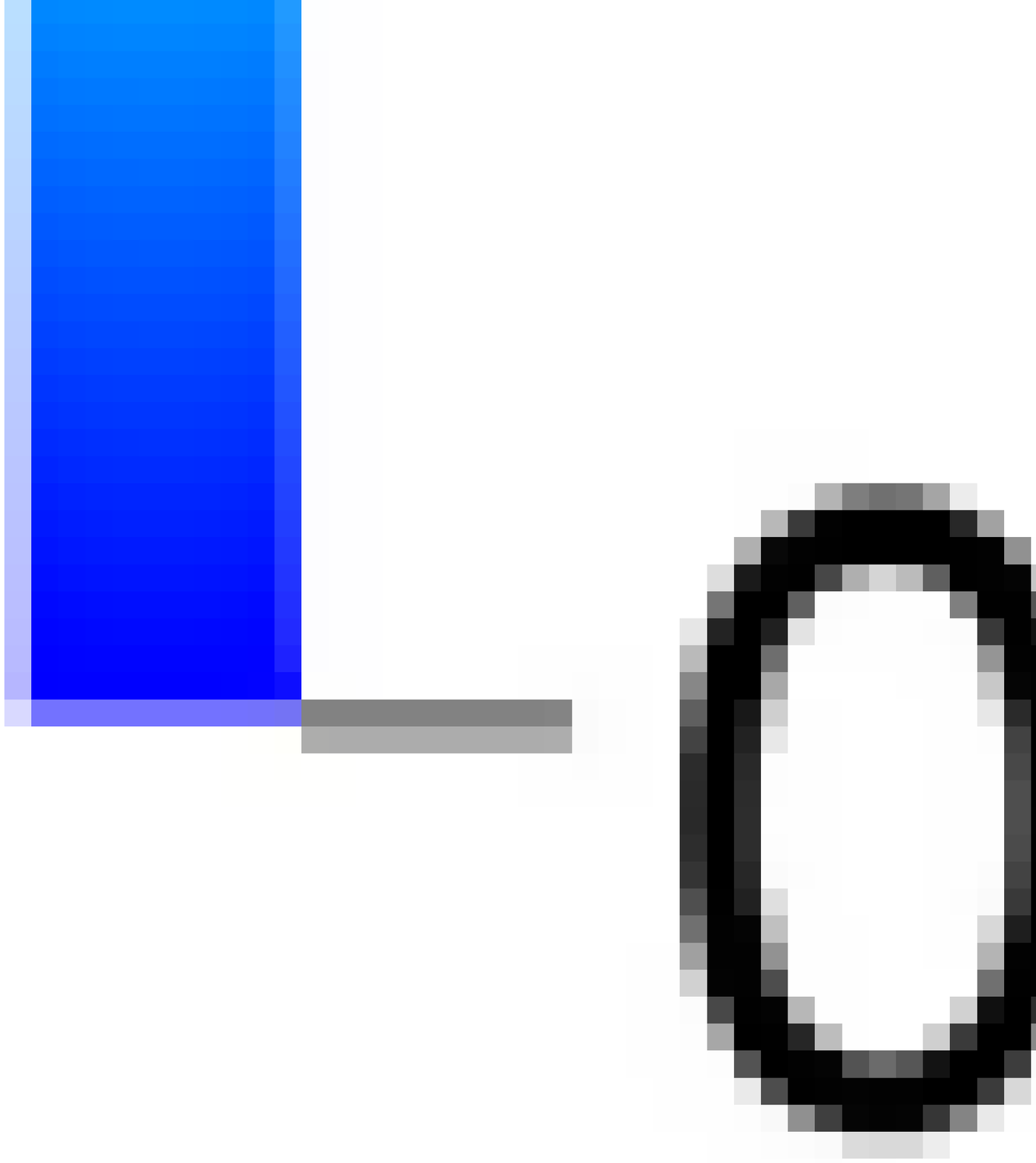}
        \end{center}
      \end{minipage}
      \begin{minipage}{0.5\hsize}
        \begin{center}
          \includegraphics[clip, width=7.5 cm]{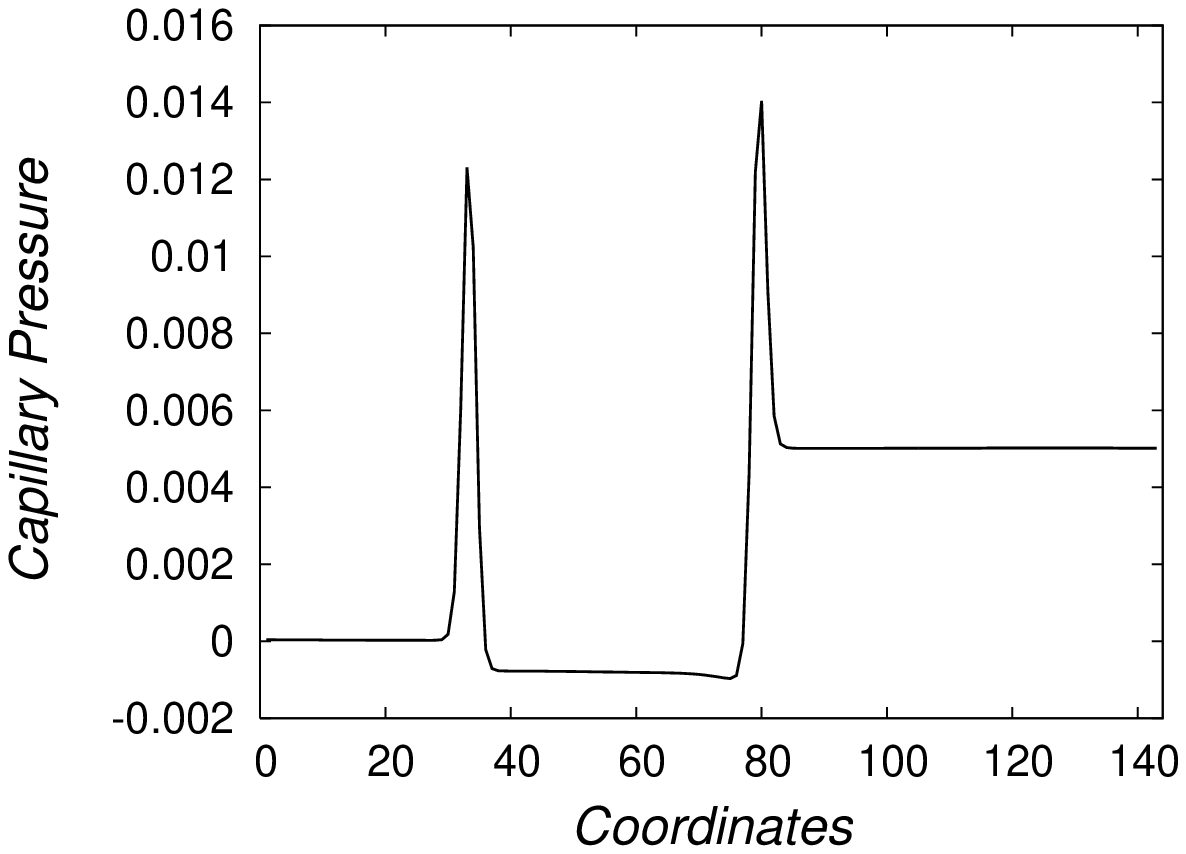}
        \end{center}
      \end{minipage}
    \end{tabular}
    \caption{Color contours of static pressure (Left) and graph of capillary pressure (Right) in the water-wet case for $Bo=1.60$ ($g=1.57e-4$) and the initial position 2. $D=8$ and $\tau_{w}=\tau_{o}=1.0$. The oil slug region where the oil density is more than half of the typical density, 0.11, is shaded. }
    \label{fig:prs_image_water-wet_tau1_pos2}
  \end{center}
\end{figure*}

\begin{figure*}[htbp]
  \begin{center}
    \begin{tabular}{c}
      \begin{minipage}{0.5\hsize}
        \begin{center}
          \includegraphics[clip, width=7.5 cm]{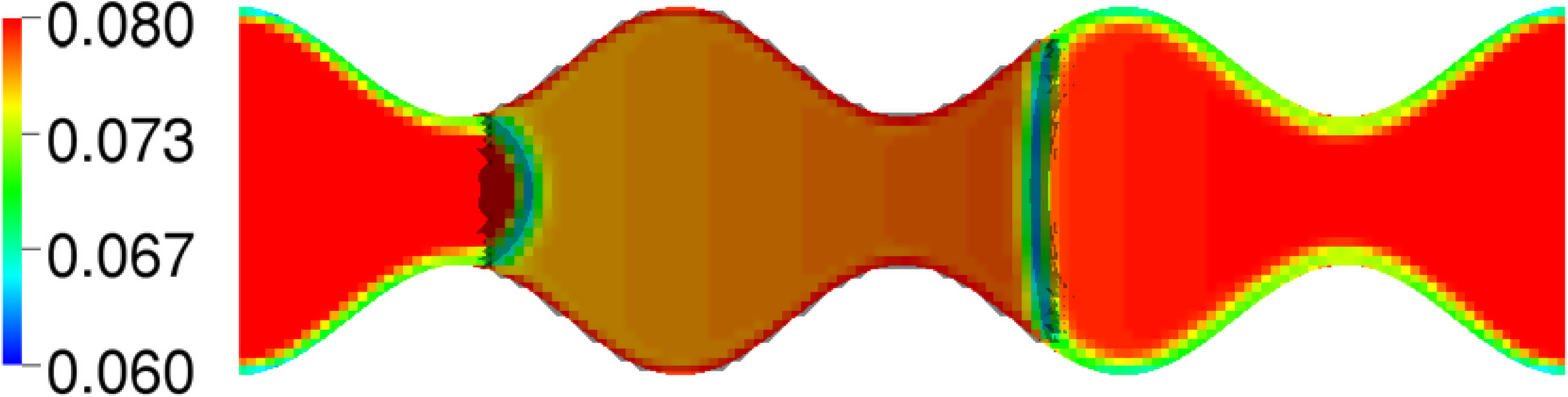}
        \end{center}
      \end{minipage}
      \begin{minipage}{0.5\hsize}
        \begin{center}
          \includegraphics[clip, width=7.5 cm]{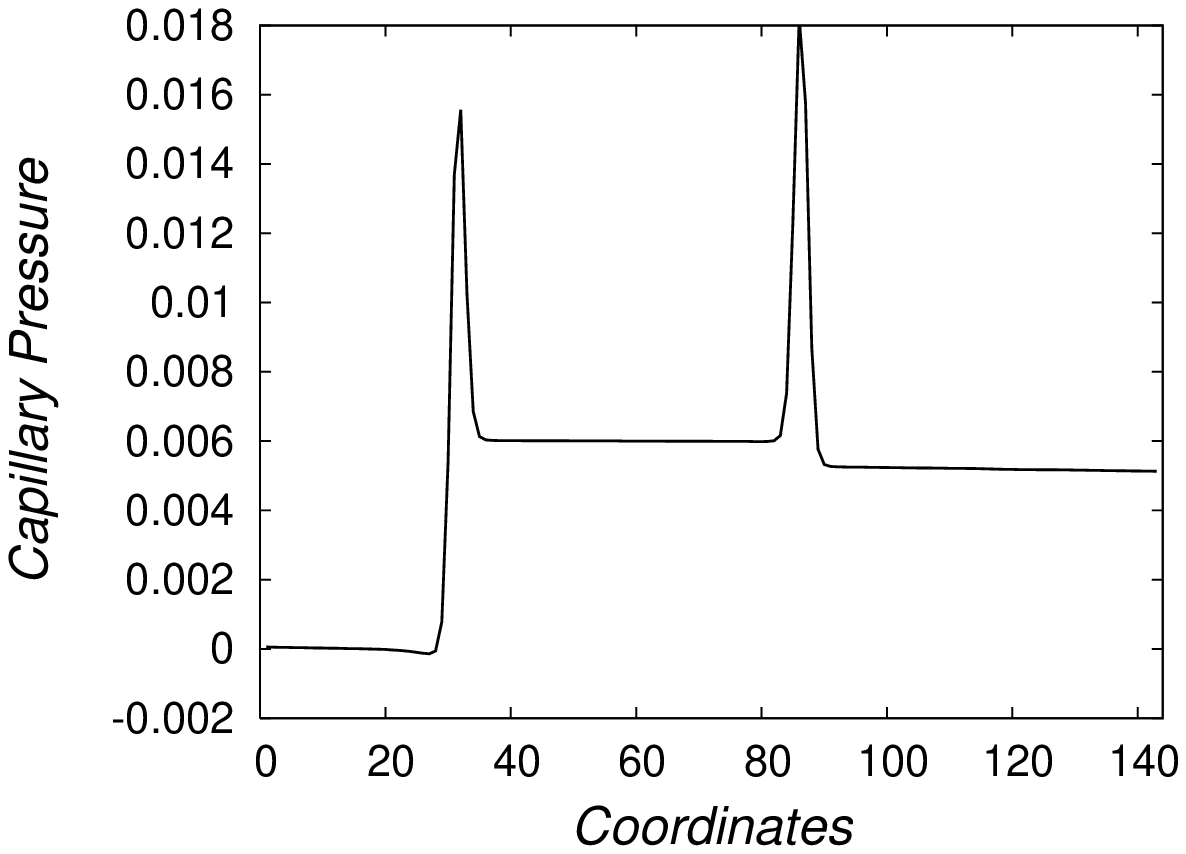}
        \end{center}
      \end{minipage}
    \end{tabular}
    \caption{Color contours of static pressure (Left) and graph of the capillary pressure (Right) for the oil-wet case for $Bo=1.62$ ($g=1.60e-4$) and the initial position 2. $D=8$ and $\tau_{w}=\tau_{o}=1.0$. The oil slug region where the oil density is more than half of the typical density, 0.11, is shaded. }
    \label{fig:prs_image_oil-wet_tau1_pos2}
  \end{center}
\end{figure*}

Although the homogeneous body force scheme induces additional density variations ($\sim 6\%$), hydrodynamic behavior of the oil slug is not much influenced by the artificial compressibility. Since the simulated critical Bond numbers agree with the incompressible analytical solution, the surface tension and capillary pressure, i.e. contact angle, are not influenced by the existence of weak compressibility in our method. In the left side of Figs. \ref{fig:prs_image_water-wet_tau1_pos2} and \ref{fig:prs_image_oil-wet_tau1_pos2}, pressure distribution in the static states of the water-wet and oil-wet cases is shown. The oil slug is shaded. Note the pronounced difference of these pressure distributions depending on the wettability. For the water-wet wall, the capillary pressure drops in the oil slug region and increases in the water region. 

Note the importance of the pore neck size for accurate determination of the critical pressure. In the capillary pressure plots shown in Figs. \ref{fig:prs_image_water-wet_tau1_pos2} and \ref{fig:prs_image_oil-wet_tau1_pos2}, there are two dominant capillary pressure jumps at the pore neck regions where the interfaces have large curvatures. The critical Bond number is mainly determined by the interface at the pore neck region. The resolution sufficient for accurate simulation was found to be 18 points across the neck.

In the water-wet case, the oil slug sometimes breaks up and leaves a small amount of oil droplet behind while moving towards its static position. It may cause noticeable critical pressure dependence upon the initial position (Table \ref{tab:crit_force_Bond_water-wet_oil-wet}). For the oil-wet case, small amounts of oil may be attracted along the wall and form a thin film layer and the breakup of oil slug is not observed.

\subsubsection{The influence of resolution}

Resolution studies using $D=6, 12$ were performed for different wettability settings and slug initial positions. The increment of $g$ is always chosen in such a way that $\Delta Bo=0.02$. The initial slug volume is always set in the same way as before. In Table. \ref{tab:Bond_res30-res15}, we show simulation results including the slug volume in the static state, and the critical Bond numbers together with the analytical critical numbers corresponding to $\Delta \theta = 0$ and $\Delta \theta \neq 0$. For each wettability condition, the results are averaged over those with four different initial slug positions. 

With the large system size $D=12$, the simulation results agree well with the analytical solutions and the relative errors are less than at most a few percent when $\Delta \theta \neq 0$. The initial condition dependence is negligible. A small breakup of the oil slug also observed in the high resolution simulation, however the volume of broken oil drop is quite small compared to the main oil slug. With $D=6$, there are only 12 lattice points across the pore neck. The critical $Bo$ numbers in several cases are overestimated relative to the analytical solutions, although the overall deviation is still within $5\%$. This can be considered quite good for such coarse simulation, especially in view of potential application to practical problems where resolution may present a bottleneck. However the accuracy of contact angle prediction is not guaranteed. Due to the lack of resolution, the numerical artifacts caused by non lattice alignment can no longer be neglected.

\subsubsection{The effects of variable viscosity}
 In this test, the component relaxation times are $\tau_{w}=1.5$ and $\tau_{o}=0.55$ for achieving the viscosity ratio of 20. The other flow conditions are identical to those in the section on wettability effects above. As we discussed early, since the critical pressure is mainly balanced by the capillary pressure, the viscous effects are not expected to be significant. However the realistic viscosity ratio is regarded to be important for other issues that are beyond the scope of the paper, such as the movement of oil slug.

Simulation results are shown in Table. \ref{table:Bo_visc}. As expected, the critical $Bo$ values are the same as those for the viscosity ratio of 1, consistent with the data presented in Figs. \ref{fig:laplace} and \ref{fig:contact_angle} above.

\section{SUMMARY $\&$ FUTURE WORK}
\label{sec:SUMMARY}
The oil slug displacement in a sinusoidal channel is simulated with an effective LBM multi-component model. The critical pressure for removing a slug is compared with available analytical solutions. The dependence of critical Bond number upon wall wettability, resolution, and fluid viscosity is systematically investigated. The simulation results agree well with theory, as shown in Table. \ref{table:sum_dev}. The main results can be summarized as follows:

\vspace{3mm}

\begin{itemize}
\setlength{\itemsep}{3mm}
\item[$\bullet$] Our multi-component LBM approach is shown effective for the quantitative study of complex multi-component fluid flows with arbitrary geometries. It is numerically accurate and stable even for very small relaxation times $\tau$ and large viscosity ratios, which are known to be challenging issues.

\item[$\bullet$] It is numerically verified that homogeneous body force can be efficiently used to replace pressure as a driving force. This reduces artificial compressibility effects without sacrificing accuracy. Numerically achieved contact angle, surface tension and critical Bond number show little dependence upon local compressibility.

\setlength{\itemsep}{3mm}
\item[$\bullet$] For accurate prediction of the critical pressure, contact angle variability caused by lattice non-alignment in arbitrary geometries can be important. This leads to hysteresis-like behavior in our studies that increases the critical pressure. Additional studies on these lattice dependence effects seem necessary.

\item[$\bullet$] Sufficient resolution for resolving flow at the pore neck region is critical for accurate prediction of the critical Bond number. Large curvature of the interfaces at the pore necks result in large capillary pressure differences.

\item[$\bullet$] Small breakup of oil slugs is sometimes observed. With the increased resolution, the impact of this effect decreases. Without sufficient resolution, however, such breakup may influence the oil slug displacement.

\item[$\bullet$] The independence of critical pressure upon viscous effects is demonstrated.

\end{itemize}

\vspace{3mm}

 Beyond the current study, we are conducting similar investigations of single component multi-phase fluid flow behavior. Using similar methods helps to address the issues of compressibility and high density ratio in consistent thermodynamics framework \cite{1994_Xiaowen}. Let us briefly discuss some potential ways to improve results such as reported here. First of all, improved models fluid interaction with the solid wall could further reduce the resolution and lattice alignment dependence of the contact angle. Also, an adaptive mesh functionality would be particularly useful for resolving the pore neck region with the obvious benefit for quantitative accuracy improvement. Last but not the least, the critical pressure should be studied with multiple oil slugs and bypass channels. For multiple slugs, the stability of solutions is an interesting topic because of its relevance for slug volume and total capillary force. For bypass channels with joined ends, when an oil slug fully blocks flow in one channel and no oil exists in the other channel, the static slug is subject to force which is almost equivalent to pressure drop in a flowing channel. Since such pressure drop is caused by friction and does depend on viscosity, the Reynolds number becomes one of critical factors in addition to the Bond number.

\section{ACKNOWLEDGMENTS}
 The authors would like to thank our colleagues in the physics group and application group of Exa corporation for important comments and discussions, and Zen Sugiyama for his great support creating the computational geometry data for this study.
\vspace{3mm}

\begin{table}[h]
  \begin{center}
  \caption{ The contact angle dependence upon the resolution across the channel height $H$ and the boundary offset relative to the lattice (in fractions of the lattice unit length). For all cases, $\tau_{w}=\tau_{o}=1.0$. The normalized wall potentials for water and oil are $\left\{ \rho_s^{w}=0, \: \rho_s^{o}=0.44 \right\}$ for the oil-wet case (Left) and $\left\{ \rho_s^{w}=0.44, \: \rho_s^{o}=0 \right\}$ for the water-wet case (Right).}
    \begin{tabular}{c}
      \begin{minipage}{0.5\hsize}
        \begin{center}
		{\tabcolsep = 0.2cm
		  \renewcommand\arraystretch{0.8}
		\begin{tabular}{|c|c|c|c|} \hline
		offset length & H=16 & H=32 & H=64 \\ \hline
        	0   & 145  & 144 & 141 \\
		0.2 & 140  & 139 & 138 \\
		0.4 & 140  & 140 & 139 \\
		0.6 & 141  & 142 & 141 \\
		0.8 & 141  & 144 & 142 \\ \hline
		    &  140-145 & 139-144 & 138-142 \\ \hline
	        \end{tabular}	
		}
        \end{center}
      \end{minipage}
      \begin{minipage}{0.5\hsize}
        \begin{center}
	    {\tabcolsep = 0.2cm
	     \renewcommand\arraystretch{0.8}
	    \begin{tabular}{|c|c|c|c|} \hline
	    offset length & H=16 & H=32 & H=64 \\ \hline
	    0   & 39.9  & 39.2 & 40.1 \\
	    0.2 & 44.1  & 43.7 & 42.6 \\
	    0.4 & 42.6  & 42.6 & 41.8 \\
	    0.6 & 40.6  & 40.4 & 40.1 \\
	    0.8 & 39.6  & 38.9 & 39.3 \\ \hline
		& 39.6-44.1 & 38.9-43.7 & 39.3-42.6 \\ \hline
	    \end{tabular}
	    }
        \end{center}
      \end{minipage}
    \end{tabular}
    \label{tab:contact_angle_pickup} 
  \end{center}
\end{table}

\begin{table}[htb]
 \begin{center}
	\caption{Targeted oil slug volume $V_{0}$ non-dimensionalized by the cubic power of characteristic length $D$. }
	{\tabcolsep = 0.1cm
         \renewcommand\arraystretch{0.8}
	\begin{tabular}{|l|c|c|} \hline
	wettability &  $V_{0}/D^3$\\ \hline
	intermediate  &  63 $\pm$ 5 \\
	water-wet     &  56 $\pm$ 5 \\
	oil-wet       &  69 $\pm$ 5  \\ \hline
	\end{tabular}
	}
	\label{table:slug_volume}
 \end{center}
\end{table}

\begin{table}[htb]
	\begin{center}
	\caption{The slug volume right before it begins to move, $V_{0}/D^3$, and critical Bond numbers from simulation and theory, as a function of wettability and slug position. Here $D=8$ and $\tau_{w}=\tau_{o}=1.0$. Data from Table. \ref{tab:contact_angle_pickup} are used in the fifth column to account for the theoretical critical $Bo$ dependence upon lattice orientation effects that result in $\Delta \theta \neq 0$. The data in the sixth column disregards this effect.}
	{\tabcolsep = 0.2cm
        \renewcommand\arraystretch{0.5}
	\begin{tabular}{|c|c|c|c|c|c|} \hline
	wettability & slug initial position          &  slug volume    &  Bo  & Bo  & Bo   \\ 
	            &                                &    $V_{0}/D^3$  &   (simulation)               &     (analysis, $\Delta \theta \neq 0$)   &   (analysis, $\Delta \theta = 0$)      \\  \hline
	water-wet   & position1  &     59.4        &    1.58 - 1.60    &  1.60         &  1.50       \\ 
	            & position2  &     57.4        &    1.60 - 1.62    &  1.62         &  1.52          \\
	            & position3  &     57.5        &    1.60 - 1.62    &  1.62         &  1.52          \\
		    & position4  &     57.1        &    1.60 - 1.62    &  1.62         &  1.52           \\ \hline
		    &            &  58.3 $\pm$ 2.3 &    1.58 - 1.62    &  1.60 -1.62         &  1.50 - 1.52      \\ \hline 
	oil-wet     & position1  &     69.5        &    1.62 - 1.64    &  1.64         &  1.50           \\
	            & position2  &     69.6        &    1.62 - 1.64    &  1.64         &  1.50            \\
		    & position3  &     69.5        &    1.62 - 1.64    &  1.64         &  1.50            \\
		    & position4  &     69.5        &    1.62 - 1.64    &  1.64         &  1.50             \\ \hline
		    &            &  69.6 $\pm$ 0.1 &  1.62 - 1.64      &  1.64         &  1.50            \\ \hline 
	\end{tabular}
	\label{tab:crit_force_Bond_water-wet_oil-wet}
	}	
	\end{center}
\end{table}

\begin{table}[h]
  \begin{center}
  \caption{The slug volume right before it begins to move, $V_{0}/D^3$, and the critical Bond number from simulation and theory, as a function of resolution and wettability. $D= \left\{ 12,6 \right\}$ and $\tau_{w}=\tau_{o}=1.0$. For each wettability condition, results for all four initial slug positions are included}
    \begin{tabular}{c}
		{\tabcolsep = 0.2cm
		  \renewcommand\arraystretch{0.5}
		\begin{tabular}{|c|l|c|c|c|c|} \hline
	simulation size	& wettability  & slug  volume  &  Bo            &    Bo                           &  Bo          \\ 
			&              &  $V_{0}/D^3$  &   (simulation) &   (analysis, $\Delta\theta$ $\neq$ $0$)    &  (analysis, $\Delta\theta$ = $0$)   \\ \hline
	$D=12$		& intermediate &   59.3 $\pm$ 0.0       &  1.56 - 1.58   &    1.58 - 1.60               &      1.58 - 1.60         \\ 
			& water-wet    &   57.0 $\pm$ 0.1       &  1.58 - 1.60   &    1.60 - 1.62               &      1.52         \\ 
			& oil-wet      &   66.4 $\pm$ 0.0       &  1.60 - 1.62   &    1.60 - 1.62               &      1.48 - 1.50         \\  \hline	
	$D=6$		& intermediate &   57.6 $\pm$ 0.7        &  1.56 - 1.58   &    1.54 - 1.56               &   1.54 - 1.56       \\ 
			& water-wet    &   56.0 $\pm$ 2.6        &  1.58 - 1.66   &    1.60 - 1.62               &   1.50 - 1.52   \\ 
			& oil-wet      &   67.5 $\pm$ 0.7        &  1.66 - 1.68   &    1.62                      &   1.50   \\  \hline
	        \end{tabular}	
		}
    \end{tabular}
    \label{tab:Bond_res30-res15}
  \end{center}
\end{table}

\begin{table}[htb]
 \begin{center}
	\caption{The slug volume right before the motion onset ,$V_{0}/D^3$ , and the critical Bond numbers from simulation and theory for $D=8$ and $\tau_{w}$=0.55, $\tau_{o}$=1.5.}
	{\tabcolsep = 0.2cm
         \renewcommand\arraystretch{0.5}
	 \begin{tabular}{|l|c|c|c|c|} \hline
	        wettability  &  slug volume        & Bo             &   Bo           &  Bo                          \\ 
		             &  $V_{0}/D^3$        & (simulation)   &   (analysis, $\Delta \theta \neq 0$)   &  (analysis, $\Delta \theta = 0$)  \\ \hline
                intermediate &  58.0 $\pm$ 0.7  &  1.56 - 1.59     &   1.56 - 1.58    &  1.56 - 1.58                   \\ 
        	water-wet    &  57.0 $\pm$ 2.2  &  1.58 - 1.62     &   1.60 - 1.62    &  1.52                        \\ 
        	oil-wet      &  69.7 $\pm$ 0.2  &  1.62 - 1.64     &   1.64           &  1.50                        \\  \hline
	\end{tabular}
	}
	\label{table:Bo_visc}
 \end{center}
\end{table}

\begin{table}[htb]
 \begin{center}
	\caption{Difference, in percentage points, between the simulated and analytical critical Bond number. The results for $D=8$ and $\tau_{w}$=0.55, $\tau_{o}$=1.5 are shown in parentheses. For all the other cases, $\tau_{w}=\tau_{o}=1$. The difference of less than a few percent is indistinguishable from the numerical error.}
	{\tabcolsep = 0.5cm
         \renewcommand\arraystretch{0.5}
	 \begin{tabular}{|l|c|c|c|c|} \hline
	        wettability  &   D=6           &    D=8                    &     D=12           \\ \hline
                intermediate &  0 - 2.6        &   0 - 1.3 (0 - 1.3)       &    0 - 2.5          \\
        	water-wet    &  0 - 3.8        &   0 - 1.3 (0 - 1.3)       &    0 - 2.5          \\
        	oil-wet      &  2.4 - 3.7      &   0 - 1.2 (0 - 1.3)       &    0 - 1.2          \\ \hline
	\end{tabular}
	}
	\label{table:sum_dev}
 \end{center}
\end{table}





\begin{thebibliography}{40}


\bibitem{1975_Morrow} Norman R. Morrow, The effects of surface roughness on contact angle with special reference to petroleum recovery, Journal of Canadian Petroleum Technology 14(1975)42

\bibitem{2013_Colosqui} Carlos E. Colosqui, Michail E. Kavousanakis, Athanasios G. Papathanasiou, Ioannis G.Kevrekidis, A mesoscopic model for microscale hydrodynamics and interfacial phenomena Slip films and contact angle hysteresis, Phys Rev E 87(2013)013302


\bibitem{1998_Chen} H.Chen, C.Teixeria, K.Molving, Int J Mod Phys C 9(1998)1281 

\bibitem{2009_Leo} Yanbing Li, R.Zhang, R.Shock, H.Chen, Prediction of vortex shedding from a circular cylinder using a volumetric Lattice-Boltzmann boundary approach, Eur.Phys.J.Special Topics 171(2009)91-97

\bibitem{2004_Yanbing} Yanbing Li, Richard shock, Raoyang Zhang, Hudong Chen, Numerical study of flow past an impulsively started cylinder by the lattice-Boltzmann method, J.Fluid Mech 519(2004)273-300

\bibitem{2006_Fan} Hongli Fan, Raoyang Zhang, Hudong Chen, Extended volumetric scheme for lattice Boltzmann models, Phys Rev E 73(2006)066708

\bibitem{2008_Michael} Michael C.Sukop, Haibo Huang, Chen Luh Lin, Milind D.Deo, Kyeongseok Oh, Jan D.Miller, Distribution of multiphase fluids in porous media: Comparison between lattice Boltzmann modeling and micro-x-ray tomography, Phys Rev E 77(2008)026710

\bibitem{2002_Manwart} C.Manwart, U.Aaltosalmi, A.Koponen, R.Hilfer, J.Timonen, Lattice-Boltzmann and finite-difference simulations for the permeability for three-dimensional porous media, Phys Rev E 66(2002)016702

\bibitem{1998_Hazlett} R.D.Hazlett, S.Y.Chen, W.E.Soll, Wettability and rate effects on immiscible displacement: Lattice Boltzmann simulation in microtomographic images of reservoir, Journal of petroleum science and engineering 20(1998)167-175

\bibitem{2006_Fredrich} J.T.Fredrich, A.A.DiGiovanni, D.R.Noble, Predicting macroscopic transport properties using microscopic image data, J Geophys Res 111(2006)B03201

\bibitem{1992_Steven} Steven Bryant, Martin Blunt, Prediction of relative permeability in simple porous media, Phys Rev A 46(1992)2004   

\bibitem{2010_Ramstad} Thomas Ramstad, P\aa l-Eric \O ren, Sting Bakke, Simulation of two-phase flow in reservoir rocks using a lattice Boltzmann method, SPE Journal 124617(2010)923-933

\bibitem{2001_Fang} Hai-ping Fang, Le-wen Fan, Zuo-wei Wang, Zhi-fang Lin, Yue-hong Qian, Studying the contact point and interface moving in a sinusoidal tube with lattice Boltzmann method, Int J Mod Phys B 15(2001)1287-1303



\bibitem{2014_Haihu} Haihu Liu, Qinjun Kang, Christopher R. Leonardi, Bruce D. Jones, Sebastian Schmieschek, Ariel Narv\'aez, John R. Willianms, Albert J. Valocchi, Jens Harting, Multiphase lattice Boltzmann simulations for porous media applications, arXiv:1404.7523

\bibitem{2007_Haibo} Haibo Huang, Daniel T.Thorne Jr., Marcel G.Schaap, Michael C.Sukop, Proposed approximation for contact angle in Shan-and-Chen-type multicomponent multiphase lattice Boltzmann models, Phys Rev E 76(2007)066701

\bibitem{2008_Haibo} Haibo Huang, Zhitao Li, Shuaishuai Liu, Xi-yun Lu, Shan-and-Chen-type multiphase lattice Boltzmann study of viscous coupling effects for two-phase flow in porous media, Int J Numer Methods Fluids 61(2009)341-354


\bibitem{2007_Andreas} Andreas G.Yiotis, John Psihogios, Michael E.Kainourgiakis, Aggelos Papaioannou, Athanassios K. Stubos, A lattice Boltzmann study of viscous coupling effects in immiscible two-phase flow in porous media, Colloids Surf A Physicochem Eng Asp 300(2007)35-49


\bibitem{1978_Mariano} Mariano A. Neira, Alkiviades C. Payatakes, Collocation solution of creeping newtonian flow through periodically constricted tubes with piecewise continuous wall profile, AlChE Journal 24(1978)43

\bibitem{1979_Mariano} Mariano A. Neira, Alkiviades C. Payatakes, Collocation solution of creeping newtonian flow through sinusoidal tubes, AlChE Journal 25(1979)725

\bibitem{1979_SOO} Soo Gun Oh, John C.Slattery, Interfacial tension required for significant displacement of residual oil, SPE Journal 19(1979)83-96

\bibitem{1993_Xiaowen} Xiaowen Shan, Hudong Chen, Lattice Boltzmann model for simulating flows with multiple phases and components, Phys Rev E 47(1993)1815

\bibitem{1995_Xiaowen} Xiaowen Shan, Gary Doolen, Multi-component lattice-Boltzmann model with interparticle interaction, arXiv:comp-gas/9503001 (1995)

\bibitem{1996_Xiaowen} Xiaowen Shan, Gary Doolen, Diffusion in a multi-component Lattice Boltzmann Equation model, arXiv:comp-gas/9605003 (1996)

\bibitem{2006_Chen} Hudong Chen, Raoyang Zhang, Ilya Staroselsky, Myung Jhon,
Recovery of full rotational invariance in lattice Boltzmann formulations for
high Knudsen number flows, Physica A 362 (1) (2006), 125

\bibitem{2006_Zhang} Raoyang Zhang, Xiaowen Shan, Hudong Chen, Efficient kinetic
method for fluid simulation beyond the Navier-Stokes equation, Phys. Rev. E, 74,
(2006) 046703

\bibitem{2006_Latt} Jonas Latt, Bastien Chopard, Lattice Boltzmann method with
regularized pre-collision distribution functions, Math. Comput. Simulat. 72 (2-6)
(2006), 165

\bibitem{Payatakes_1973} Alkiviades C. Payatakes, Chi Tien, Raffi M. Turian, A New Model for Granular Porous Media: Part I. Model Formulation, AIChE Journal 19(1), (1973), 58-67

\bibitem{2006_Xiaowen_JFM} Xiaowen Shan, Xue-Feng Yuan, Hudong Chen, Kinetic theory representation of hydrodynamics: a way beyond the Navier-Stokes equation, J.Fluid Mech 550(2006)413-441

\bibitem{1997_Chen} Hudong Chen, Chris Teixeira, Kim Molving, Digital physics approach to computational fluid dynamics: some basic theoretical features, Int.J.Mod.Phys.C 8(1997)675

\bibitem{1992_Qian} Y.Qian, D.d'Humi\'eres, P.Lallemand, Lattice BGK models for Navier-stokes equation, Europhys.Lett 17(1992)479

\bibitem{1991_Chen} S.Chen, H.Chen, D.Martnez, W.Matthaeus, Lattice Boltzmann model for simulation of magnetohydrodynamics, Phys Rev Lett 67(1991)3776

\bibitem{1992_Chen} H.Chen, S.Chen, W.H.Matthaeus, Recovery of the Navier-Stokes equations using a lattice-gas Boltzmann method, Phys Rev A 45(1992)R5339


\bibitem{1954_Bhantnagar} Bhatnagar P.L, Gross E, M.Krook, A model for collisions in gases I. Small amplitude processes in charged and neutral one-component systems, Phys Rev 94(1954)511-525

\bibitem{2012_Qli} Q.Li, K.H.Luo, X.J.Li, Forcing scheme in pseudopotential lattice Boltzmann model for multiphase flows, Phys Rev E 86(2012)016709

\bibitem{2008_Xiaowen} Xiaowen Shan, pressure tensor calculation in a class of nonideal gas lattice Boltzmann model, Phys Rev E 77(2008)066702

\bibitem{2012_Connington} Kevin Connington, Taehun Lee, A review of spurious currents in the lattice Boltzmann method for multiphase flows, Journal of mechanical science and technology 26(2012)3857-3863 

\bibitem{2008_Wagner} Alexander J. Wagner, The origin of spurious velocities in lattice Boltzmann, Int J Mod Phys B 17(2003)193

\bibitem{2006_Xiaowen} Xiaowen Shan, Analysis and reduction of the spurious current in a class of multiphase lattice Boltzmann models, Phys Rev E 77(2006)047701

\bibitem{1994_Xiaowen} Xiaowen Shan, Hudong Chen, Simulation of non-ideal gases and liquid-gas phase transitions by lattice Boltzmann equation, Phys Rev E 49(1994)2941












\end{thebibliography}
\end{document}